\documentclass{elsartL}
\usepackage{graphicx}
\usepackage{amssymb}
\usepackage{amsmath}
\usepackage{physics}
\usepackage{mathrsfs}
\usepackage{mathtools}
\usepackage{rotating}
\usepackage{appendix}
\newcommand{\RNum}[1]{\uppercase\expandafter{\romannumeral #1\relax}}
\begin{document}

\begin{frontmatter}
\title{A discussion on two celebrated examples of application of linear regression}
\author{Evangelos Matsinos}

\begin{abstract}
This work aims at providing some (further) perspective into the analysis of two well-known datasets in terms of the methods of linear regression. The first set has been taken from Hubble's 1929 investigation 
of the relation between the distances of galaxies and their recessional velocities. The second set relates to Galton's family data on human stature, collected for the purposes of his 1889 book on natural 
inheritance.\\
\noindent {\it PACS 2010: 07.05.Kf, 02.60.Pn}
\end{abstract}
\end{frontmatter}

\section{\label{sec:Introduction}Introduction}

Regression methods have been indispensable tools in the analysis of observations for over two centuries. The first description of linear regression may be traced to one of Legendre's books \cite{Legendre1805}, 
published in 1805. It seems that Gauss had already developed and employed the technique before Legendre's publication, but had considered it ``trivial'' and unworthy of any special attention, e.g., see 
Ref.~\cite{Priceonomics}.

The purpose of regression analysis is to provide optimal solutions (and predictions obtained thereof) to overdetermined systems, i.e., to situations where the available amount of information (i.e., the number 
of the input datapoints) exceeds the number of parameters being involved in the modelling of the processes yielding the observations. The simplest of the regression models involves the replacement of the 
entirety of the available information (e.g., of all lifespans of incandescent light bulbs, produced by a specific manufacturer) by one `constant' (i.e., by the mean product lifetime) and by one uncertainty 
(e.g., by the `root mean square' of the distribution of the lifetimes of the tested products): information is thus imparted to the potential customer that a purchased bulb of that specific brand and type is 
expected to withstand, say, $8\,000$ h of operation with an estimated uncertainty of, say, $500$ h. Consequently, it will be very surprising if such a bulb eventually exceeds $12\,000$ h of operation, and 
(in case that the underlying distribution of the product lifetime is symmetrical about the mean value) equally surprising if it becomes dysfunctional before it reaches $4\,000$ h of operation~\footnote{In 
practice, however, the latter tends to be more probable than the former, though Statistics is hardly to blame for that mishap!}. The next-to-simplest modelling involves a straight line~\footnote{Although a 
constant may be thought of as a straight line without slope, such a `semantic narrowing' in the definition of the straight line will be avoided in this work.} (and two adjustable parameters, intercept and 
slope).

Numerous authors have occupied themselves with the details of the methods of linear regression (ordinary, weighted, multiple, etc.). The task of listing even the most influential of these contributions would 
inevitably result in a very long reference list, and, in all probability, even such a list would be incomplete. Therefore, I have decided to simply quote a few relevant books from my own collection 
\cite{Carroll1988,Maddala1992,Graybill1994,Hastie2016,Darlington2019} and rather refer the interested reader to the works cited therein, as well as to two articles which discuss the results of the application 
of some standard methods of linear regression to simulated and experimentally acquired datasets, see Refs.~\cite{Cantrell2008,Mikkonen2019}.

This work discusses details of the application of linear regression to two celebrated datasets, relating to different branches of Science and obtained in the distant past: the first dataset (Section 
\ref{sec:Hubble}) is taken from Astronomy, from the 1929 paper by Edwin Powell Hubble (1889-1953) on the relation between the distance $r$ and the radial velocity $v$ of galaxies \cite{Hubble1929}, which 
became known as `Hubble's law' later on; the second dataset (section \ref{sec:Galton}) is taken from Anthropometry, from the 1889 (the year Hubble was born) book by Francis Galton (1822-1911) on natural 
inheritance \cite{Galton1889}. As the present paper grew longer than I initially thought, I decided to abandon my plan to include herein a discussion on two additional datasets.

All important details of the methods of linear regression which are applied to the two aforementioned datasets may be found in Appendix \ref{App:AppA}. That appendix serves as a short guide to obtaining the 
formalism which is relevant to the various methods of linear regression, including the determination of the optimal values of the parameters and of their uncertainties, as well as of any predictions based 
on the results of the fits (including all correlations among the fit parameters). It is recommended to the reader to start with the material in the appendices, in order to become familiar, if not with the 
`theoretical' details about the various analysis options, at least with the definitions.

For the sake of brevity, a number of acronyms will be used in this paper:
\begin{itemize}
\item PDF will stand for `Probability Density Function';
\item CDF will stand for `Cumulative Distribution Function';
\item DoF will stand for `degree of freedom';
\item NDF will stand for `number of DoFs';
\item WLR will stand for `weighted linear regression' (`weighted least-squares optimisation'); and
\item rms will stand for `root mean square' (square root of the unbiased variance of a dataset).
\end{itemize}
Finally, pc will stand for parsec, defined as the distance at which the length of one astronomical unit ($1 {\rm AU} \coloneqq 149\,597\,870.7$ km), representative of the Sun-Earth distance, subtends an 
angle of one arcsecond: $1~{\rm pc} \coloneqq 1~{\rm AU} \, (\tan \theta)^{-1}$, where $\theta \coloneqq \pi / 1\,296\,000$ rad; $1~{\rm pc} \approx 3.085678 \cdot 10^{13}$ km or (after using the former 
unit of length for cosmic distances, namely the light year) $1~{\rm pc} \approx 3.262~{\rm ly}$.

\section{\label{sec:Data}The datasets}

\subsection{\label{sec:Hubble}Hubble's dataset}

\subsubsection{\label{sec:HubbleTheBeginnings}The beginnings}

The story behind Hubble's determination of the famous constant, which has been named after him, started over $100$ years before Hubble was born: Clotho's spindle started spinning when Edward Pigott (1753-1825), 
an English astronomer, developed a fondness for stars of variable brightness. Pigott's family moved to York in 1781, where Pigott met John Goodricke (1764-1786), a deaf-mute with unparalleled visual acuity 
and sensitivity. Working in a private observatory built by Pigott's father, Pigott and Goodricke embarked on a project aiming at exploring the properties of the most promising variable stars, taken from 
Pigott's precompiled list \cite{French2012}. It did not take long before Goodricke's skills and attention to detail paid off: one year into the project, he announced to the Royal Society the discovery of the 
periodical variation of the brightness of the star Algol ($\beta$ Persei) \cite{Goodricke1783}. Apart from proposing a mechanism to account for the effect (and attributing it to the transits of Algol's dimmer 
binary companion), Goodricke also examined the variations of brightness of several other stars, including $\eta$ Aquilae (A) and $\delta$ Cephei (A) \cite{Goodricke1786}, which became textbook examples of 
the Cepheid family of variable stars~\footnote{Cepheid variable stars pulsate radially, varying in both diameter and temperature \cite{Shapley1914}.}, featuring very regular `sharply-up-gradually-down' light 
curves, dissimilar to the one obtained from Algol; evidently, the variation of brightness of these stars could not be attributed to stellar eclipses. Goodricke's contributions to Astronomy did not go 
unnoticed: at the age of 21, he was elected a Fellow of the Royal Society, a few days before his sad passing.

\subsubsection{\label{sec:HubbleTheHCO}The Harvard College Observatory}

Over the course of the next century, many more Cepheid variable stars were detected. With the invention of photography, the acquisition of accurate brightness data was boosted. The Harvard College Observatory 
became the hub of the developments in Observational Astronomy, and its new director in 1877, Edward Charles Pickering (1846-1919), launched an ambitious programme aiming at photographing and cataloguing all 
observable stars. Whether due to budget issues (e.g., see Ref.~\cite{Singh2005}, p.~205) or to the poor quality of the work of Pickering's male assistants \cite{Rossiter1980}, over $80$ College-educated women 
were recruited, to work at the Observatory as \emph{human computers}, i.e., as skilled employees, comparing photographic plates, processing astronomical data, and contributing to Astronomy for the first time 
after Hypatia of Alexandria and Aglaonice of Thessaly.

The \emph{computers} excelled in their assignments. For instance, Annie Jump Cannon (1863-1941), who was deaf after she turned twenty, conducted the main work in the development of the Harvard Classification 
Scheme of stars, the first attempt towards a categorisation of the stars in terms of their surface temperature and spectral type. In an article entitled `Woman making index of $100\,000$ stars for a catalogue', 
the Danville Morning News reported on 10 February 1913 \cite{Danville1913}: ``When the work was new she could analyze at the rate of $1\,000$ stars in three years. Now she analyzes $5\,000$ stars in one month, 
$200$ stars an hour~\footnote{This is an erratum: evidently, the author of the article intended to write `day', not `hour'.}. On Jan.~1 she had examined about $65\,000$, which means that about two-fifths of 
the work is completed.''

Without doubt, Pickering's research programme reached its climax with another \emph{computer}, Henrietta Swan Leavitt (1868-1921). Leavitt's methodical work was destined to make an indelible impact on 
Astronomy (and Cosmology). Having been given the task of studying variable stars in the Small and Large Magellanic Clouds (SMC and LMC, respectively), Leavitt catalogued $1\,777$ such objects between 1903 
and 1908 \cite{Leavitt1908}. The decisive moment came when she focussed her attention on a small subset of these objects, i.e., on the Cepheid variable stars in the SMC (see Table \RNum{6} of 
Ref.~\cite{Leavitt1908}); on p.~107 of her report, she inconspicuously commented: ``It is worthy of notice that in Table \RNum{6} the brighter variables have the longer period.''

Following her intuition in subsequent years, Leavitt examined the relation between the period of the variation of the apparent magnitude (which is associated with the apparent brightness) of the Cepheid 
variable stars in the SMC and the extrema of the apparent magnitude for each such object. Her decision to restrict her analysis to the SMC turned out to be of pivotal importance, because the selected $25$ 
(i.e., the $16$ from Table \RNum{6} of Ref.~\cite{Leavitt1908} and the $9$ which were added to her database between 1908 and 1913) celestial bodies were, in practical terms, equidistant from the 
Earth~\footnote{It is currently known that the SMC is a dwarf galaxy (with a typical diameter of about $5.78$ kpc), gravitationally attracted to the Milky Way, and separated from the Earth by $60.5(7.0)$ 
kpc \cite{NED}.}.

After plotting the period-brightness~\footnote{I shall use this short term when referring to the datapoints corresponding to the period and to the apparent magnitude of variable stars; other authors prefer 
the term `period-luminosity'.} datapoints of the $25$ Cepheid variable stars on a linear-log plot (logarithmic scale on the $x$ axis, linear scale on the $y$ axis), Leavitt~\footnote{The fact that Pickering, 
rather than Leavitt, signed the 1912 report is baffling. The report starts with the sentence: ``The following statement regarding the periods of $25$ variable stars in the Small Magellanic Cloud has been 
prepared by Miss Leavitt.'' Although Pickering's motive for signing Leavitt's report is a mystery to me, I decided to include him in the author list of that report, yet as second author.} obtained the 
``remarkable'' result shown in Fig.~2 of Ref.~\cite{Leavitt1912}: two nearly parallel straight lines, modelling the apparent-magnitude maxima and minima of these stars, separated by about $1.2$ mag 
(difference between the corresponding intercepts); I shall use `mag' to denote the unit of apparent magnitude (which, of course, is not a physical unit). Two remarks from that report serve as a prelude to 
the significance of Leavitt's research:
\begin{itemize}
\item ``\dots there is a simple relation between the brightness of the variables and their periods'' and
\item ``Since the variables are probably at nearly the same distance from the Earth, their periods are apparently associated with their actual emission of light, as determined by their mass, density, and 
surface brightness.''
\end{itemize}
The consequences of Leavitt's findings were beyond reckoning. The measurement of the period of a Cepheid variable star, whichever neighbourhood of the Universe that specific celestial body occupies, could 
lead to an estimate for its apparent brightness $b_{\rm SMC}$ which that body would have if it had been confined within the SMC. By comparing the body's measured apparent brightness $b$ to $b_{\rm SMC}$, one 
could obtain an estimate for the distance between the specific Cepheid variable star and the Earth, expressed (of course) in terms of the SMC-Earth distance. As the distances of all Cepheid variable stars 
would be expressed as multiples of the SMC-Earth distance, one would need to determine only one such distance (e.g., using the stellar-parallax method) in order that all distances become known. Within one 
year of Leavitt's breakthrough, Ejnar Hertzsprung (1873-1967) provided a solution, however flawed~\footnote{On p.~204 of Hertzsprung's article \cite{Hertzsprung1913}, one reads: ``\dots so wird die 
entsprehende visuelle Sterngr{\"o}{\ss}e gleich $13^m.0$. Diese {\"U}berlegung f{\"u}hrt also zu einer Parallaxe $p$ der kleinen Magellanschen Wolke, welche durch $5 \log p = -7.3 - 13.0 = -20.3$ gegeben ist. 
Man erh{\"a}lt $p = 0^{\prime \prime}.0001$, einem Abstand von etwa $3\,000$ Lichtjahren entsprechend.'' This part is translated as follows: ``\dots in which case the corresponding apparent magnitude is 
equal to $13^m.0$. This consideration leads to a parallax of the Small Magellanic Cloud, which is given by the relation $5 \log_{10} p = -7.3 - 13.0 = -20.3$. One obtains $p = 0^{\prime \prime}.0001$, which 
corresponds to a distance of about $3\,000$ light years.'' It is a puzzle to me how Hertzsprung obtained this result. Expressed in arcseconds, the parallax would be equal to about $8.710 \cdot 10^{-5}$, 
yielding a distance of about $3.543 \cdot 10^{20}$ m or about $37\,448$ ly. However that may be, it is known today that the SMC-Earth distance is significantly larger than both aforementioned values.}, to 
this problem \cite{Hertzsprung1913}. In spite of the serious underestimation of the cosmic distances when using Hertzsprung's results, Harlow Shapley (1885-1972), then at the Mount Wilson Observatory, 
elaborated on Hertzsprung's method and obtained a better calibration of the cosmic distances on the basis of the apparent brightness of the Cepheid variable stars \cite{SawyerHogg1965}. The yardstick to 
measure cosmic distances had been invented. Or hardly so?

Looking at Leavitt's period-brightness data from the comfort of my armchair (next to a fast desktop, running present-day software) over one century after she (arduously) collected the data of the $25$ 
Cepheid variable stars, I notice that the values of Pearson's correlation coefficient between the logarithm of the period and both apparent-magnitude extrema are very close to $-1$: $-0.966$ in case of the 
apparent-magnitude maxima and $-0.962$ in case of the apparent-magnitude minima, both calling for linear regression. To advance to the modelling of these data, one first needs to establish a method to assess 
the uncertainties relevant to the apparent magnitude (there is no mention of uncertainties in Leavitt's reports), and, to be able to select a reasonable model, one first needs to revisit the procedure by 
which the apparent-magnitude data are obtained.

The apparent brightness $b$ of a star is linked to the star's luminosity $L$, which is the absolute measure of the power emitted by the star in the form of electromagnetic radiation of all frequencies, via 
the formula
\begin{equation} \label{eq:EQ001}
b = \frac{L}{4 \pi r^2} \, \, \, ,
\end{equation}
where $r$ stands for the distance between the points of emission and absorption, i.e., for the star-Earth distance. From the measurement of the apparent brightness, one obtains the apparent magnitude of the 
star by using the relation
\begin{equation} \label{eq:EQ002}
m = - \frac{5}{2} \log_{10} \frac{b}{b_f} \, \, \, ,
\end{equation}
where the quantity $b_f$ is the reference apparent brightness (i.e., the apparent brightness which is to be associated with the apparent magnitude of $0$). It must be borne in mind that \emph{smaller} values 
of the apparent magnitude are indicative of \emph{higher} apparent brightness: the apparent magnitude of Venus lies between $-4.92$ and $-2.98$, of Sirius ($\alpha$ Canis Majoris) about $-1.46$, and of Polaris 
($\alpha$ Ursae Minoris) about $1.98$; these three celestial bodies have been listed in order of decreasing apparent brightness.

Using Eq.~(\ref{eq:EQ002}), one obtains the uncertainty $\delta m$ from the uncertainty of the apparent brightness $\delta b$, as follows:
\begin{equation} \label{eq:EQ003}
\delta m = \frac{5}{2 \, \ln 10} \frac{\delta b}{b} \, \, \, .
\end{equation}
Although $m$ depends on the choice of the reference apparent brightness $b_f$, $\delta m$ is independent of that choice. It is reasonable to assume that the uncertainties of the apparent brightness are 
proportional; under this assumption, the uncertainties $\delta m$ are constant. Without doubt, the matter is more complex than it has been presented, yet the discussion above suffices for `the sake of argument' 
purposes. 

Following the aforementioned train of thought, constant working uncertainties ($\delta m = 0.2$ mag) were assigned to all apparent-magnitude data of Ref.~\cite{Leavitt1912}. It must be emphasised that, 
provided that all assigned uncertainties are equal (and non-zero), the exact value of $\delta m$ has no bearing whatsoever on the results.

\begin{figure}
\begin{center}
\includegraphics [width=15.5cm] {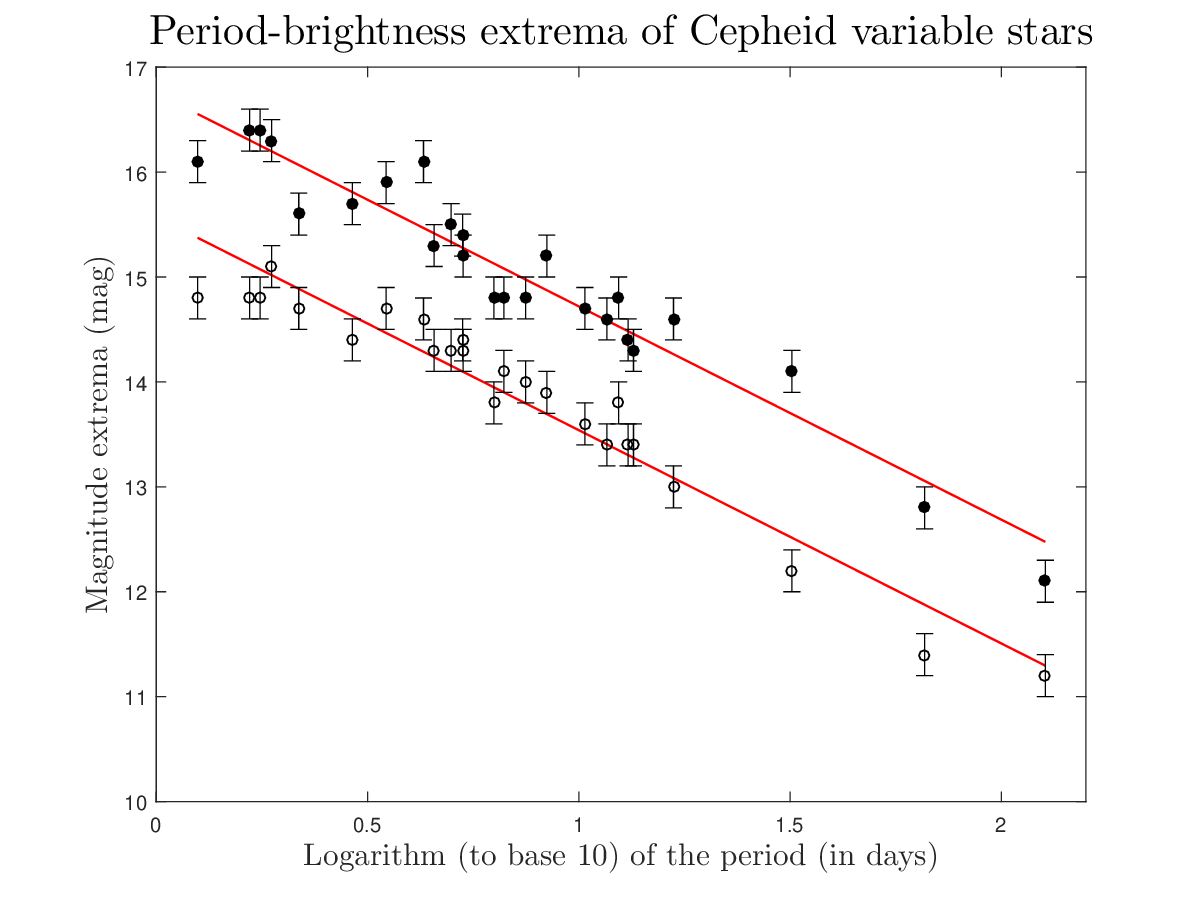}
\caption{\label{fig:LeavittCepheids}The results of the joint fit to the period-brightness data of the $25$ Cepheid variable stars of Ref.~\cite{Leavitt1912} (Table \RNum{6} therein). Filled circles: 
apparent-magnitude minima; unfilled circles: corresponding apparent-magnitude maxima. The value of the slope appears positive in Fig.~2 of Ref.~\cite{Leavitt1912}, because the vertical axis is shown inverted 
in that figure, see also caption of Table \ref{tab:Leavitt}.}
\vspace{0.5cm}
\end{center}
\end{figure}

The separate linear fits to the data (compare the second and third columns of Table \ref{tab:Leavitt}) suggest that there is nothing against the hypothesis that the two datasets be accounted for on the basis 
of a common slope parameter. The third row of Table \ref{tab:Leavitt} contains the results of the joint fit to the data, using as parameters: the intercept $p_0$ of the straight line which models the 
apparent-magnitude maxima, the common slope parameter $p_1$, and the (positive) difference $\Delta p_0$ between the intercepts of the two straight lines which model the apparent-magnitude extrema. The results 
of the joint fit to Leavitt's period-brightness data are shown in Fig.~\ref{fig:LeavittCepheids}.

\vspace{0.5cm}
\begin{table}[h!]
{\bf \caption{\label{tab:Leavitt}}}The results, contained in the second and third columns of this table, correspond to the separate fits to the maxima and the minima of the apparent magnitude, see Table 
\RNum{6} of Ref.~\cite{Leavitt1912}; the results of the joint (three-parameter) fit (featuring a common slope parameter) are shown in the last column. The value of the slope appears positive in Fig.~2 of 
Ref.~\cite{Leavitt1912}, because the vertical axis is shown inverted in that figure, evidently to conform with the common-sense expectation that brightness maxima appear `above' brightness minima.
\vspace{0.25cm}
\begin{center}
\begin{tabular}{|l|c|c|c|}
\hline
 & Fit to maxima & Fit to minima & Joint fit\\
\hline
\hline
$N$ & $25$ & $25$ & $50$\\
NDF & $23$ & $23$ & $47$\\
$\chi^2_{\rm min}$ & $40.89$ & $47.88$ & $88.84$\\
$\hat{p}_0$ (mag) & $15.56$ & $16.77$ & $15.573$\\
$\hat{p}_1$ & $-2.02$ & $-2.05$ & $-2.033$\\
$\Delta \hat{p}_0$ (mag) & - & - & $1.180$\\
$\delta \hat{p}_0$ (mag) & $0.11$ & $0.12$ & $0.088$\\
$\delta \hat{p}_1$ & $0.11$ & $0.12$ & $0.082$\\
$\delta (\Delta \hat{p}_0)$ (mag) & - & - & $0.078$\\
t-multiplier & $1.022217$ & $1.022217$ & $1.010752$ \\
Corrected $\delta \hat{p}_0$ (mag) & $0.11$ & $0.12$ & $0.089$\\
Corrected $\delta \hat{p}_1$ & $0.11$ & $0.12$ & $0.083$\\
Corrected $\delta (\Delta \hat{p}_0)$ (mag) & - & - & $0.079$\\
\hline
\end{tabular}
\end{center}
\vspace{0.5cm}
\end{table}

I left two comments for the very end of the story regarding Leavitt's invaluable contributions to Science. First, Leavitt too was suffering from severe hearing problems by the time she joined Pickering's 
team in 1903. Second, like Goodricke $135$ years earlier, she did not live long enough to enjoy the fruits of her labour.

\subsubsection{\label{sec:HubbleContributions}Hubble and Company}

It was about time Hubble, a U.S.~American from the state of Missouri, came on stage. Hubble had been offered a job at the Mount Wilson Observatory already in 1916, but arrived with a delay of about three 
years after he answered the call of duty and embarked on the task of defending the United Kingdom against the Quadruple Alliance. When Hubble finally arrived at the Observatory, the $100$-inch Hooker Telescope 
had been completed and was about to yield unprecedented observations of the Cosmos for the next few decades. Shapley was still at the Observatory, at the peak of his scientific productivity \cite{SawyerHogg1965}. 
That was a time when a paradigm shift was under way in the human understanding of the Cosmos, with the introduction of General Relativity a few years earlier and the gradual emergence of Quantum Mechanics. 
Astronomers were striving to provide answers to questions about the nature of the `nebulae', which were detected at an increasing rate with the help of the modern telescopes, and the size of the Universe; 
a relevant question was whether or not the nebulae were part of the Milky Way. There was no shortage of arguments for a few years, as the case frequently is when reliable evidence is scarce. Intending to 
resolve the matter, the National Academy of Sciences proposed a public debate between representatives of the two sides in the first half of 1920 \cite{Trimble1995}; this event became known as the `Great 
Debate', though - for the obvious reasons - I would rather replace `Debate' with `Divide'. The two representatives, Shapley (for those who propounded that the nebulae were within the Milky Way) and Heber 
Doust Curtis (1872-1942) (for those who surmised that the nebulae were independent galaxies), were different in nearly all ways possible \cite{Singh2005}, not only regarding their views on the matter at hand: 
they had different personalities, different ways of expressing themselves when providing explanations to others, different skills when addressing the general public, and - the icing on the cake - different 
social background. On 27 April 1920 (the day after the Great Debate), there was no clear winner: the unquestionable winner was announced four years later, and his name was Hubble.

Hubble's resolution of the question of the Great Debate was based on a somewhat serendipitous discovery: on 4 October 1923, he took a long exposure of the Andromeda Nebula (M31) \cite{Singh2005} and noticed 
peculiar `spots' on the photographic plate. Following up on the issue the next night, he found out that two of the spots were novae, whereas the third one turned out to be a Cepheid variable star, the first 
one to be detected beyond the two Magellanic Clouds. He evaluated the object's distance and found it several times larger than the typical dimensions of the Milky Way, suggesting that the Andromeda Nebula 
was an independent galaxy, well beyond our own. Several other Cepheid variable stars were detected in M31 shortly afterwards (also by Hubble), confirming the earlier result \cite{CepheidsM31}. Following the 
scientific code of conduct, Hubble broke the news to Shapley (who had moved to Cambridge, Massachusetts, to assume the position of the Director of the Harvard College Observatory) in early 1924. After 
reading Hubble's letter, Shapley is said to have commented to one of his doctoral students: ``Here is the letter that destroyed my universe,'' see Ref.~\cite{Trimble1995}, p.~1142. That was Hubble's first 
significant contribution to Observational Astronomy. He then turned his attention to a related problem.

In the early 1910s, Vesto Melvin Slipher (1875-1969) became the first astronomer to measure the radial velocity of a nebula. Utilising the Doppler shift, he determined the radial velocity of the Andromeda 
Galaxy (which, as aforementioned, still had the status of a nebula at those times) relative to the Solar System, and found out that it approaches the Milky Way at a speed of $300$ km/s, see first paper 
of Refs.~\cite{Slipher}. Within a few years, Slipher had enlarged his database to a total of $25$ nebulae (see Table 1 of the last paper of Refs.~\cite{Slipher}), $21$ of which were undoubtedly receding; in 
addition, four of the receding nebulae had radial velocities in excess of $1\,000$ km/s (which was enormous in comparison with the typical relative velocities of stars - about $30$ km/s - in the Milky Way). 
Provided that there is no preferred direction in the motion of these objects, the probability that at most $4$ nebulae (in a set of $25$) either recede from or approach the Milky Way is equal to about 
$9.11 \cdot 10^{-4}$. This probability suggests a statistically significant departure~\footnote{In this study, the threshold of statistical significance, which is frequently denoted in the literature as 
$\alpha$, is assumed to be equal to $1.00 \cdot 10^{-2}$. This threshold is regarded by most statisticians as signifying the outset of statistical significance. A usual choice in several branches of Science 
is $\alpha=5.00 \cdot 10^{-2}$ (which most statisticians associate with `probable statistical significance').} from the null hypothesis that the distribution of the radial velocities of the nebulae is 
symmetrical about $0$. Slipher's results posed the inevitable question: why do so many more nebulae \emph{retreat} from the Milky Way?

In the early 1920s, Alexander Alexandrovich Friedmann (1888-1925) derived (from General Relativity) the equations which became known in Cosmology as `Friedmann's equations' \cite{Friedmann1922,Friedmann1924}: 
one of the solutions of these equations predicted an expanding Universe (see also Ref.~\cite{Matsinos2017}, Chapters 10.3.4 and 10.8.3). A few years later (and working independently), Georges Henri Joseph 
{\'E}douard Lema$\hat{\rm \i}$tre (1894-1966), the father of the Big-Bang Theory with his seminal 1931 paper \cite{Lemaitre1931}, also came up with the same equations and attributed the recession of the 
nebulae to the expansion of space \cite{Lemaitre1927}. It is far less known (or, perhaps, intentionally overlooked) that what many call `Hubble's law' can be found in Lema$\hat{\rm \i}$tre's 1927 paper 
\cite{Lemaitre1927}: the ratio $v/r$ (where, as mentioned in Section \ref{sec:Introduction}, $r$ represents the distance of a nebula from the Earth and $v$ its radial velocity), which in subsequent years 
became known as Hubble's constant~\footnote{Regarding Hubble's constant, a caustic comment I usually make is that it is neither Hubble's nor a constant (given its dependence on the cosmological time); $H_0$ 
is the value of the Hubble parameter at the current cosmological epoch.} $H_0$, was actually evaluated by Lema$\hat{\rm \i}$tre to $625$ km/s Mpc$^{-1}$ (see fifth line from the top of p.~56 of Lema$\hat{\rm \i}$tre's 
paper \cite{Lemaitre1927}), not far from Hubble's two estimates a few years later. Unfortunately for Lema$\hat{\rm \i}$tre, his paper had been published in a journal of very limited dissemination. In 
addition, Lema$\hat{\rm \i}$tre's efforts in 1927, to draw attention to his work, were not successful \cite{ORaifeartaigh2020}. The consequence was that the famous relation between the distance and the 
radial velocity at which the galaxies recede from the Milky Way became known as `Hubble's law', instead of the fairer `Lema$\hat{\rm \i}$tre-Hubble law' (see also Ref.~\cite{Matsinos2017}, Chapter 10.3.1).

\begin{figure}
\begin{center}
\includegraphics [width=15.5cm] {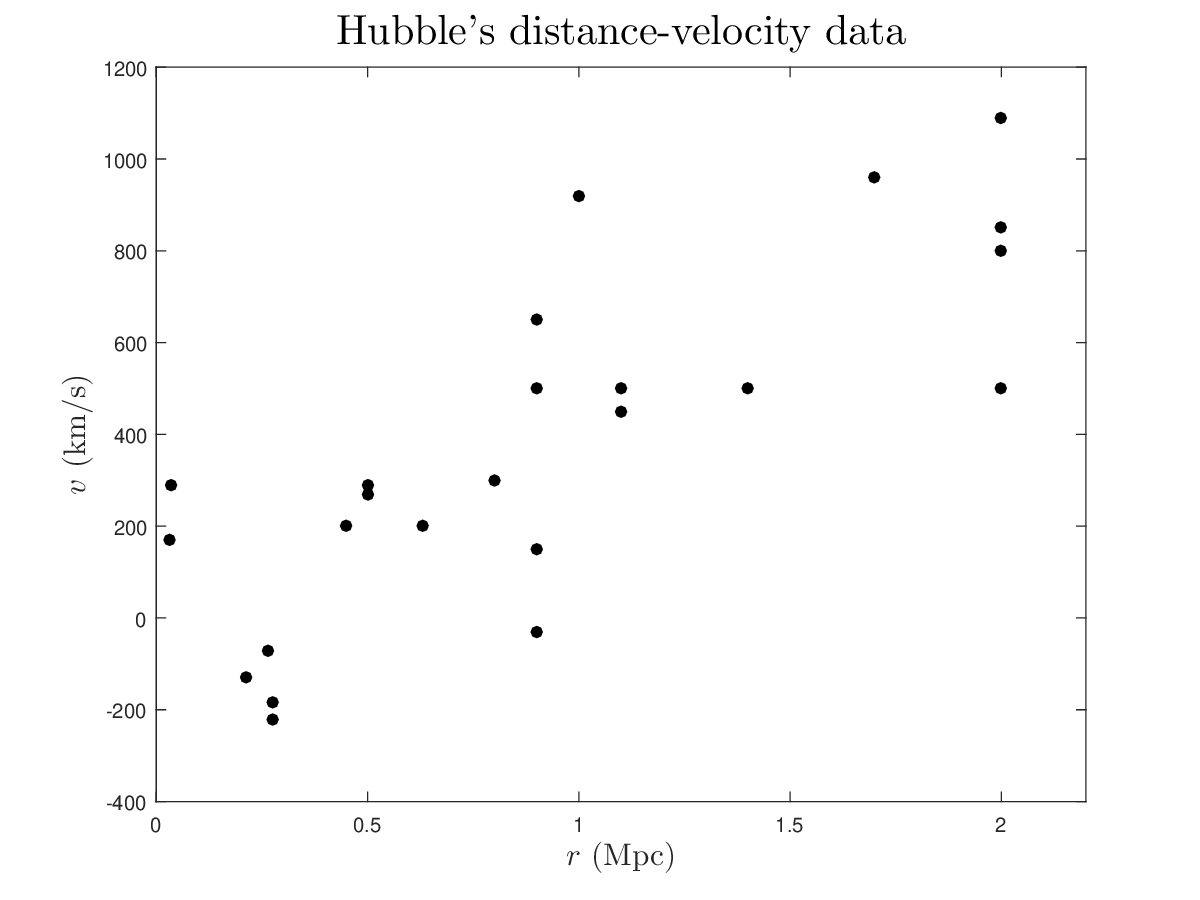}
\caption{\label{fig:Hubble}The pairs of measurements of the distance and of the radial velocity of the nebulae on which Hubble based his main analysis in Ref.~\cite{Hubble1929}.}
\vspace{0.5cm}
\end{center}
\end{figure}

Concerning the data on which Hubble based his $H_0$ determination, I have heard two Cosmologists remark in their seminars: ``Hubble's data seem to be all over the place'' and ``I do not know what Hubble did 
in his paper.'' I find both comments lamentable (though, regrettably, I did not raise an objection on those two occasions) and shall next address both issues. Available to Hubble in his 1929 paper 
\cite{Hubble1929}, were pairs of measurements of the distance $r$ and of the radial velocity $v$ of nebulae. To account for the observations, Hubble employed the linear model
\begin{equation} \label{eq:EQ003a}
v = K r + \Delta v \, \, \, ,
\end{equation}
where $K$ was assumed to be a constant (to be identified with $H_0$) and the correction
\begin{equation} \label{eq:EQ004}
\Delta v = X \, cos \delta \, \cos \alpha + Y \, \cos \delta \, \sin \alpha + Z \, \sin \delta
\end{equation}
compensated for the motion of the Solar System (i.e., for the motion of the observer); the velocities $X$, $Y$, and $Z$ were parameters, to be obtained from the optimisation of the description of the 
observations. The angles $\alpha$ and $\delta$ in Eq.~(\ref{eq:EQ004}) denote the right ascension and the declination of each observed object, respectively. Hubble's $(r,v)$ datapoints, originating from $24$ 
``nebulae whose distances have been estimated from stars involved or from mean luminosities in a cluster,'' see Table 1 of Ref.~\cite{Hubble1929}, are shown in Fig.~\ref{fig:Hubble}.

Was Hubble wrong when suggesting that the distances of the nebulae and their radial velocities are linearly related? Do the data spread ``all over the place?'' The answer to both questions is negative. 
Pearson's correlation coefficient for the data shown in Fig.~\ref{fig:Hubble} is equal to $\rho \approx 0.790$. Given that the resulting value of the $t$-statistic is about $6.036$, one obtains the p-value 
of about $4.48 \cdot 10^{-6}$ (two-tailed). Therefore, there is ample evidence to reject the null hypothesis that the two quantities, $r$ and $v$, are not correlated: equivalently, Hubble's data do \emph{not} 
spread all over the place.

Using Hubble's datapoints, I obtained the following results for $K$ (i.e., $H_0$), $X$, $Y$, and $Z$: $465(56)$ km/s Mpc$^{-1}$, $-69(79)$ km/s, $235(152)$ km/s, and $-200(82)$ km/s, respectively. Hubble 
quotes slightly different values and sizeably smaller uncertainties in Ref.~\cite{Hubble1929} (in all probability, he employed constraints in his fit). His values read as follows: $465(50)$ km/s Mpc$^{-1}$, 
$-65(50)$ km/s, $226(95)$ km/s, and $-195(40)$ km/s, respectively. After enhancing his database, Hubble updated his results two years later \cite{Hubble1931}; nevertheless, his $H_0$ value remained enormous 
(in fact, the $H_0$ estimate of 1931 is even larger than the 1929 result!), suggesting that the typical age of the Universe (obtained from $H_0^{-1}$) was smaller than the age of the oldest rocks found on 
the Earth (established via radiometric dating), which in turn provides an estimate for the temporal interval which elapsed since the Earth's crust was formed. Hubble's $H_0$ result stood for over two decades, 
and so did the age-of-the-Universe paradox (also known as ``timescale difficulty'').

It was Wilhelm Heinrich Walter Baade (1893-1960), another `user' of the $100$-inch Hooker Telescope at the Mount Wilson Observatory (as well as of the $200$-inch Hale Telescope at the Palomar Observatory 
in the late 1940s), who came up with an explanation for the paradox. Hubble's results had been flawed, because his estimates for the distances of the galaxies had been erroneous. It had not occurred to him 
that the Cepheid variable stars basically come in two types: the classical ones, which are young luminous stars, and the older ones which are fainter, e.g., see Ref.~\cite{ATNF}, figure in Section `Calculating 
distances using Cepheids'. The distance calibration had involved the \emph{second} type, whereas Hubble had been processing data pertaining to \emph{classical} Cepheid variable stars. The consequence was that 
the distances of the galaxies had seriously been underestimated (to be \emph{so} luminous, the observed star had to be closer to the Milky Way). By how much had Hubble underestimated the present-day distances? 
It turns out that Hubble's distances were between $6.8$ and $68.7$~\% of the present-day estimates (available from the NASA/IPAC Extragalactic Database \cite{NED}), the median being equal to about $15.3$~\%. 
(I find it very surprising that no one validated Hubble's determination of the distances of the galaxies for such a long time.) Within a few years of Baade's discovery, his doctoral student, Allan Rex Sandage 
(1926-2010), came up with the first realistic estimate for $H_0$ and for the age of the Universe in his 1958 paper \cite{Sandage1958}; in the abstract of that work, one reads: ``\dots This gives 
$H \approx 75$ km/s Mpc$^{-1}$ or $H^{-1} \approx 13 \times 10^9$ years \dots''

I had been wondering for some time what Hubble's $H_0$ result would have been, if present-day data for the galaxies, which are mentioned in Ref.~\cite{Hubble1929}, had been available to him. To this end, I 
used the latest information from NED \cite{NED} for all $45$ galaxies mentioned in Ref.~\cite{Hubble1929}, i.e., for the then selected $24$ (whose data Hubble considered `reliable'), as well as for the $21$ 
which, though mentioned in Hubble's paper, had not been fully measured by 1929. Information about the right ascension, declination, and radial velocity can be found directly in the main web page of these 
galaxies in NED, whereas all available distance estimates may be downloaded using a tab therein. It does not take long to realise that the variability of the distance estimates of these galaxies is sizeable 
even today, nearly one century after Hubble's analysis. In any case, I obtained the median, the mean, and the rms of the distribution of \emph{all} distance estimates (no exclusion of results) for each of 
these galaxies, and finally constructed the input (to the optimisation software) datafile using the means and the rms values; alternatively, one could replace the means with the medians.

The measured radial velocities were corrected for the motion of the Solar System, using updated information regarding the `peculiar' velocity \cite{Schoenrich2010} and the rotational motion of the Local 
Standard of Rest (LSR) \cite{Reid2014,Reid2016}, a coordinate system obtained after averaging the velocities of stars in our neighbourhood. The three components of the velocity of the Solar System relative 
to the centre of the Milky Way will be denoted as $(U, V, W)$: in the coordinate system which follows the right-hand rule (to my great surprise, the use of left-handed coordinate systems is not uncommon in 
Observational Astronomy!), $U$ is positive towards the galactic centre, $V$ is positive in the direction of the galactic rotation, and $W$ is positive towards the North Galactic Pole. The velocity of the 
Solar System relative to the centre of the Milky Way is obtained by adding the peculiar velocity of the Sun (i.e., the velocity relative to the LSR) to the velocity of the LSR relative to the centre of the 
Milky Way. The velocity components $U$ and $W$ (and their uncertainties) were taken from Ref.~\cite{Schoenrich2010}, whereas $V$ (and its uncertainty) was fixed from Ref.~\cite{Reid2014}; the $V$ result of 
Ref.~\cite{Reid2014} is compatible with the estimate for the rotational velocity of the LSR which is recommended in Ref.~\cite{Reid2016}. For the transformation from equatorial $(\alpha,\delta)$ to galactic 
(latitude $b$ and longitude $l$ of each celestial object) angles, Ref.~\cite{Carroll2007} was followed, see p.~900 therein: the radial velocities of Ref.~\cite{NED} were corrected by adding to the measured 
velocity the term $\Delta v \, = \, U \, \cos b \, \cos l \, + \, V \, \cos b \, \sin l \, + \, W \, \sin b$.

\begin{figure}
\begin{center}
\includegraphics [width=15.5cm] {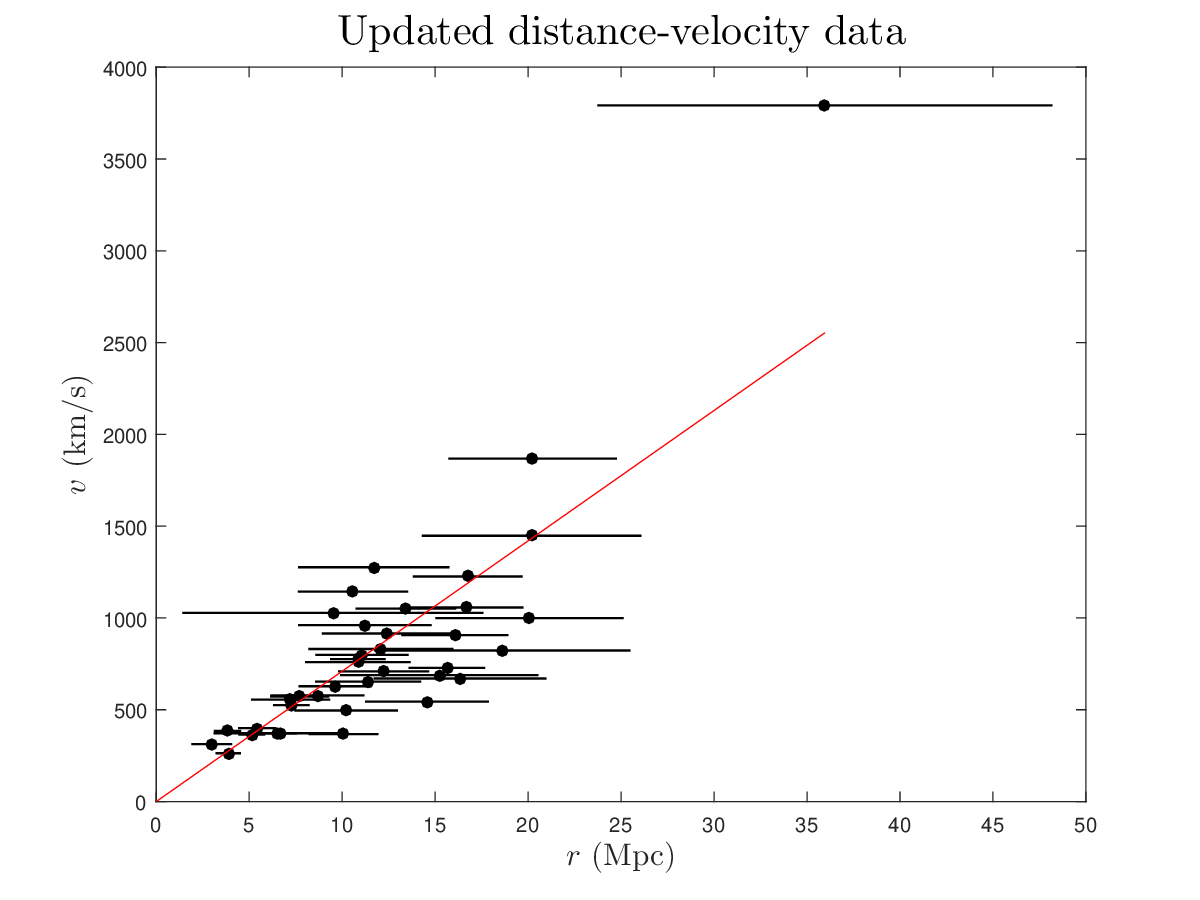}
\caption{\label{fig:HubbleUpdated}The pairs of measurements of the distance and of the radial velocity (corrected for the motion of the Solar System) of the galaxies mentioned in Ref.~\cite{Hubble1929}; the 
values correspond to present-day information, obtained from the NASA/IPAC Extragalactic Database \cite{NED}. Seven outliers have been removed from the database (see text). The uncertainties of the radial 
velocity have been taken into account, but are too small to be displayed.}
\vspace{0.5cm}
\end{center}
\end{figure}

A few words regarding the treatment of a few outliers in the dataset of the $45$ galaxies, mentioned in Ref.~\cite{Hubble1929}, are due.
\begin{itemize}
\item After its radial velocity was corrected, NGC 3031 appears to recede from the Milky Way at a speed of $74.9(3.8)$ km/s, whereas its distance suggested that it should recede at about $250$ km/s.
\item Six galaxies (SMC, LMC, NGC 6822, NGC 598, NGC 221, and NGC 224) are separated from the Milky Way by less than $1$ Mpc: the proximity of these galaxies to the Milky Way results in irregular radial 
velocities. The (corrected) recessional velocities of four of these galaxies are negative, implying that they approach the Milky Way (only LMC and NGC 6822 recede). Regarding the treatment of these six 
galaxies, one may either replace them with one representative object, situated at a typical distance (weighted average) of $0.055(21)$ Mpc and having a typical velocity (weighted average) of $(-5 \pm 24)$ 
km/s, or exclude them from the database.
\end{itemize}
The results of the WLRs (without intercept) following the two approaches of Appendix \ref{App:AppA4}, for four ways of treatment of the aforementioned outliers, are listed in Table \ref{tab:HubbleUpdated}; 
the input datapoints (after all seven outliers are removed from the database) and the fitted straight line in case of the EV$_2$ method are shown in Fig.~\ref{fig:HubbleUpdated}. The crux of the matter is 
that, had present-day data been available to Hubble for his 1929 paper, he probably would have ended up with $H_0=(71.0 \pm 2.9)$ km/s Mpc$^{-1}$ (or with a similar result). The Particle Data Group (PDG) 
recommend: $H_0=67.4(0.5)$ km/s Mpc$^{-1}$, obtained from a plethora of precise astronomical data \cite{PDG2022}.

\vspace{0.5cm}
\begin{table}[h!]
{\bf \caption{\label{tab:HubbleUpdated}}}The results of the linear fit (without intercept, see Appendix \ref{App:AppA4}) to the pairs of measurements of the distance and of the radial velocity (corrected for 
the motion of the Solar System) of the galaxies mentioned in Ref.~\cite{Hubble1929}; the input $(r,v)$ datapoints correspond to present-day information, available from the NASA/IPAC Extragalactic Database 
\cite{NED}. Results are quoted for four ways of treatment of the seven outliers, i.e., of the $(r,v)$ measurements corresponding to NGC 3031, as well as to the six galaxies closest to the Milky Way (see text). 
The t-multipliers, corresponding to $1 \sigma$ effects in the normal distribution, have been applied to the quoted uncertainties.
\vspace{0.25cm}
\begin{center}
\begin{tabular}{|l|c|c|c|c|}
\hline
Method & $\chi^2_{\rm min}$ & $\chi^2_{\rm min}$/NDF & $\hat{p}_1 \equiv H_0$ & $\delta \hat{p}_1 \equiv \delta H_0$\\
 & & & (km/s Mpc$^{-1}$) & (km/s Mpc$^{-1}$)\\
\hline
\hline
\multicolumn{5}{|c|}{All galaxies: $45$ $(r,v)$ datapoints}\\
\hline
Quadratic summation (EV$_2$) & $723.12$ & $16.43$ & $82$ & $14$\\
Linear summation & $494.14$ & $11.23$ & $75.4$ & $9.9$\\
\hline
\multicolumn{5}{|c|}{All but NGC 3031: $44$ $(r,v)$ datapoints}\\
\hline
Quadratic summation (EV$_2$) & $690.61$ & $16.06$ & $83$ & $14$\\
Linear summation & $468.56$ & $10.90$ & $77$ & $10$\\
\hline
\multicolumn{5}{|c|}{No NGC 3031, six closest galaxies replaced by one: $39$ $(r,v)$ datapoints}\\
\hline
Quadratic summation (EV$_2$) & $48.22$ & $1.27$ & $71.0$ & $2.9$\\
Linear summation & $45.60$ & $1.20$ & $71.0$ & $2.9$\\
\hline
\multicolumn{5}{|c|}{All outliers removed: $38$ $(r,v)$ datapoints}\\
\hline
Quadratic summation (EV$_2$) & $48.08$ & $1.30$ & $71.0$ & $2.9$\\
Linear summation & $45.47$ & $1.23$ & $71.0$ & $2.9$\\
\hline
\end{tabular}
\end{center}
\vspace{0.5cm}
\end{table}

\clearpage
\subsection{\label{sec:Galton}Of sweet peas, moths, and men: Galton's family data on human stature}

Galton was an English multidisciplinarian, with important contributions to subjects in a variety of branches of Science, ranging from Anthropology and Biology to Statistics. Before I started writing this 
article, I did not know that, in addition, Galton was among the very first to introduce weather maps in order to obtain forecasts, in particular to predict storms. Nor was I aware of the fact that he had 
been one of the pioneers of Forensic Science: in 1892, he published a book on the analysis of fingerprints as the means of identifying individuals \cite{Galton1892}. In this paper, I shall focus on his 
1889 book \cite{Galton1889}, which touches upon the subject of natural inheritance.

In his book, Galton set out to explore three questions (see pp.~1-2). I let him give a summary of these issues, copying a few sentences from his book.
\begin{itemize}
\item ``The large do not always beget the large, nor the small the small, and yet the observed proportions between the large and the small in each degree of size and in every quality, hardly varies from one 
generation to another.'' Galton suggests that corresponding PDFs of the human height ``hardly vary from one generation to another,'' though (so far as human stature is concerned) the children are not close 
replicas of their parents.
\item ``The second problem regards the average share contributed to the personal features of the offspring by each ancestor severally. Though one half of every child may be said to be derived from either 
parent, yet he may receive a heritage from a distant progenitor that neither of his parents possessed as \emph{personal} characteristic. Therefore the child does not on the average receive so much as one 
half of his \emph{personal} qualities from each parent, but something less than a half.'' At this point, the laws of Mendelian inheritance, which (presumably unbeknownst to him) had been developed by the 
time Galton started writing his book, come to my mind. (Such inherited features from distant progenitors, categorised as `latent elements', are addressed in Chapter \RNum{11} of Galton's book.)
\item ``The last of the problems that I need mention now, concerns the nearness of kinship in different degrees. We are all agreed that a brother is nearer akin than a nephew, and a nephew than a cousin, 
and so on, but how much nearer are they in the precise language of numerical statement?'' Here, Galton expresses interest in the quantification of the degree of consanguinity.
\end{itemize}

I shall next address a few issues relevant to Galton's family data on human stature (`Galton's family data' henceforth). Before skimming through Galton's book, I was under the impression that Galton's assistants 
had been responsible for all measurements of the height of the individuals on which the study \cite{Galton1889} was based. This turned out to be untrue. Galton discusses the data-acquisition process in 
Chapter \RNum{6} of his book, starting on p.~71: ``I had to collect all my data for myself, as nothing existed, so far as I know, that would satisfy even my primary requirement. This was to obtain records of 
at least two successive generations of some population of considerable size. They must have lived under conditions that were of a usual kind, and in which no great varieties of nurture were to be found.'' 
Although, according to his first sentence, Galton ``had to collect all his data for himself,'' the data was actually collected `through the offer of prizes' by individuals who were inexperienced with acquiring 
measurements for scientific purposes (pp.~74-75). These individuals simply responded to Galton's request (p.~72): ``Mr.~Francis Galton offers $500$\emph{l.} (i.e., \pounds$500$) in prizes to those British 
Subjects resident in the United Kingdom who shall furnish him before May 15, 1884, with the best Extracts from their own Family Records.'' Although the conditions regarding the use of any submitted data were 
thoroughly addressed in the advertisement (pp.~72-74), those pertaining to the data acquisition itself were not! A certain amount of disappointment overtook me as I read the following passage in Galton's book 
(p.~78): ``In many cases there remains considerable doubt whether the measurement refers to the height with the shoes on or off; not a few of the entries are, I fear, only estimates, and the heights are 
commonly given only to the nearest inch.'' He did not mention that this was mostly the result of his glaring failure to specify the process by which the measurements of the height should have been acquired. 
Nevertheless, Galton's spirits must have had a miraculous recovery as, a few lines on, he shifted his mind-set towards the thesis that ``a fair share of these returns are undoubtfully careful and thoroughly 
trustworthy, and as there is no sign or suspicion of bias, I have reason to place confidence in the values of the Means that are derived from them.'' How can there be ``no sign or suspicion of bias,'' if the 
fraction of measurements, which had been made with the shoes on, remains unknown? Without doubt, the inclusion of such measurements into the database is bound to introduce bias, in particular (for the sake 
of example) if most of the fathers walked around the house (and let their heights be measured) in shoes, whereas most of the other family members preferred to walk (and let their heights be measured) 
barefoot.

Be that as it may, Galton's records contain rows corresponding to different families: the essential information comprises the parental heights, as well as the ordered heights of all filial descendants, first 
of the sons, followed by those of the daughters. To be able to analyse the data in his time~\footnote{It was Carl Pearson (1857-1936), one of Galton's doctoral students, who brought Statistics to another 
level.} (using one independent variable), Galton decided to combine the two parental heights into one measure of `parental tallness'. As the arithmetical mean of the two heights did not appeal to him~\footnote{On 
p.~6 of his book, Galton writes: ``A serious complexity due to sexual differences seems to await us at every step when investigating the problems of heredity \dots The artifice is never to deal with female 
measures as they are observed, but always to employ their male equivalents in the place of them. I transmute all the observations of females before taking them in hand, and thenceforward am able to deal with 
them on equal terms with the observed male values.''}, he came up with a scheme of `transmuting females into male equivalents' by increasing the heights of all females in his study (i.e., of the mothers and 
of the daughters) by $8$~\%. When I first had a look at Galton's family data, this factor appeared arbitrary to me, but (as I shall explain shortly) it emerges from the data sample which was available to him. 
After `transmuting the mothers into father equivalents', Galton created a hypothetical parent (by averaging the two values), a `mid-parent' in his own words (p.~87): ``The word `Mid-Parent' \dots expresses 
an ideal person of composite sex, whose stature is half way between the stature of the father and the transmuted stature of the mother.''

Galton finally presented the data in tabular form, see Table 11 of his book, p.~208; all heights are expressed in inches ($1^{\prime\prime} = 2.54$ cm). Given in this table are the frequency distributions 
of the filial height in nine histogram bins of the mid-parent's height ranging between $63.5^{\prime\prime}$ and $72.5^{\prime\prime}$, with one overflow (mid-parent's height in excess of $72.5^{\prime\prime}$) 
and one underflow (mid-parent's height below $63.5^{\prime\prime}$) bin. As I was not content with this table, I started investigating whether Galton's original data had appeared elsewhere.

This was how I came across Ref.~\cite{Hanley2004} and found out that its author too, James A.~Hanley, had been in pursuit of the original data over two decades ago: on p.~238 of his paper, Hanley writes: 
``And so, in 2000, I began my search for Galton's `untransmuted' data \dots I also hoped that the children would still be found with their families, that is, before they were marshalled into what Galton 
called `filial' arrays.'' Hanley then continues to recount his efforts to localise Galton's records, which were finally `unearthed' in 2001, in possession of the University College London. I can almost 
sense his contentment as I read his text: ``The data were exactly what I had wished, in a single notebook, family by family, with sons and daughters identified, and with all female heights untransmuted. 
Because of the frail condition of the notebook, photocopying was not permitted \dots In February 2003, I requested and obtained permission to digitally photograph the material.'' Digital photographs of the 
eight relevant pages of Galton's notebook are now, as a result of Hanley's commendable efforts, available online \cite{GaltonOriginal}.

I copied the values, as they appear in the digital photos, onto an Excel file (the constant value of $60^{\prime\prime}$ had been subtracted from all entries in Galton's notebook) and edited a few family 
records as follows:
\begin{itemize}
\item Family 70: one daughter, described as `tall', was removed;
\item Family 76: one son (between the heights of $68^{\prime\prime}$ and $67^{\prime\prime}$) and a second daughter, both described as `medium', were removed;
\item Family 92: one daughter, described as `tall', was removed;
\item Family 95: one daughter, described as `deformed', was removed;
\item Family 118: the fourth son, described as `tallish', and all five daughters, described as `tallish' (twice), `medium' (twice), and `shortish', were removed;
\item Family 119: the first three daughters, described as `tall' (twice) and `medium', were removed;
\item Family 131: three daughters, described as `medium' and `short' (twice), were removed;
\item Family 144: the third son, described as `deformed', was removed;
\item Family 158: one daughter (between the heights of $62^{\prime\prime}$ and $61^{\prime\prime}$), described as `short', was removed;
\item Family 159: the first three daughters, described as `very tall' and `tall' (twice), were removed;
\item Family 173: the last son, described as `short', was removed;
\item Family 176: one son (between the heights of $68.5^{\prime\prime}$ and $66.5^{\prime\prime}$), described as `medium', and the last daughter, described as `idiotic', were removed;
\item Family 177: two daughters (between the two reported values), described as `tall', were removed; and
\item Family 189: the second son and one daughter (between the two reported values), both described as `middle', were removed.
\end{itemize}
No measurement of height is mentioned in the aforementioned cases of deleted entries. I thus came up with $205$ rows (families) and a total of $934$ children, $481$ sons and $453$ daughters. I cannot explain 
why Galton writes on p.~77 of his book: ``I was able to extract \dots the statures of $205$ couples of parents, with those of an aggregate of $930$ of their adult children of both sexes.'' Nor can I understand 
why the filial multiplicities in the eleven (i.e., $9+2$) bins of the mid-parent's height in his Table 11 (p.~208) sum up to $928$.

\subsubsection{\label{sec:GaltonCorrelations}Correlations between the parental and filial heights}

It is interesting to compare the values of Pearson's correlation coefficient between the sets of heights corresponding to all combinations of parental and filial types. I am aware of three methods taking on 
the problem of unequal dimensions of the input datasets ($205$ pairs of parents, $481$ sons, $453$ daughters).
\begin{itemize}
\item Method A: The analysis is restricted to the families with either exactly one son \emph{or} exactly one daughter. (Of course, families with exactly one son \emph{and} exactly one daughter contribute to 
both cases.) As the heights of the sons and of the daughters will be analysed separately, the correspondence between each \emph{pair} of parents and the relevant filial descendant (male and/or female, as the 
case might be) is one-to-one (bijective). In the accepted data, there are $42$ families with exactly one son (the multiplicity of the daughters in these families is irrelevant) and $60$ with exactly one 
daughter (the multiplicity of the sons in these families is irrelevant).
\item Method B: Within each family, the mean fraternal and sororal heights are evaluated. (Of course, families with no sons or with no daughters do not contribute to the corresponding dataset.) Again, a 
bijective correspondence between each pair of parents and the filial descendants is created (all sons and all daughters within a family are replaced by two persons - one male, another female - with stature 
corresponding to the aforementioned means). In the accepted data, there are $179$ families with at least one son and $176$ with at least one daughter.
\item Method C: The obvious commonality between the descendants within the same family is ignored, and the two parental heights are assigned to all filial descendants of each family. The correspondence is 
multi-valued: each pair of parents is linked to all of their filial descendants. 
\end{itemize}
I do not expect the introduction of bias into the analysis in cases of methods A and B.

The values of Pearson's correlation coefficient, obtained when following the aforementioned methods, are given in Table \ref{tab:Correlations}. Regardless of the method, the correlation of the heights of 
both filial types with the paternal height comes out stronger.

\vspace{0.5cm}
\begin{table}[h!]
{\bf \caption{\label{tab:Correlations}}}Pearson's correlation coefficient between the heights corresponding to all combinations of parental and filial types.
\vspace{0.25cm}
\begin{center}
\begin{tabular}{|l|c|c|}
\hline
 & Paternal height & Maternal height\\
\hline
\hline
\multicolumn{3}{|c|}{Method A}\\
\hline
Son's height & $0.642$ & $0.426$\\
Daughter's height & $0.489$ & $0.442$\\
\hline
\multicolumn{3}{|c|}{Method B}\\
\hline
Sons' mean height & $0.520$ & $0.364$\\
Daughters' mean height & $0.494$ & $0.367$\\
\hline
\multicolumn{3}{|c|}{Method C}\\
\hline
Sons' heights & $0.393$ & $0.323$\\
Daughters' heights & $0.429$ & $0.305$\\
\hline
\end{tabular}
\end{center}
\vspace{0.5cm}
\end{table}

\subsubsection{\label{sec:GaltonDistributions}Distributions of human height}

Table \ref{tab:GaltonHeights} contains the approximate values of the mean and of the rms of the distributions of parental and filial heights corresponding to Galton's 205 families. The apparent similarities 
of these statistical measures for the male (fathers and sons) and for the female (mothers and daughters) subjects, which the contents of the table suggest, require further examination.

\vspace{0.5cm}
\begin{table}[h!]
{\bf \caption{\label{tab:GaltonHeights}}}Approximate mean and rms values of the distributions of parental and filial heights corresponding to Galton's 205 families, containing acceptable entries of $481$ 
sons and $453$ daughters. The ratio of the mean parental heights is equal to about $1.083$, i.e., close to the factor of $1.08$ which Galton employed in order to transform the heights of the females into 
those of the `male equivalents'. In this work, there is no need for such a transformation.
\vspace{0.25cm}
\begin{center}
\begin{tabular}{|l|c|c|}
\hline
Family member & Mean value & rms\\
\hline
\hline
Father & $69.316^{\prime\prime}$ & $2.647^{\prime\prime}$\\
Mother & $64.002^{\prime\prime}$ & $2.333^{\prime\prime}$\\
Son & $69.234^{\prime\prime}$ & $2.624^{\prime\prime}$\\
Daughter & $64.103^{\prime\prime}$ & $2.356^{\prime\prime}$\\
\hline
\end{tabular}
\end{center}
\vspace{0.5cm}
\end{table}

The four distributions of the parental and the filial heights will next be submitted to a number of tests. Implementations of the algorithms relevant to these tests are available in the form of classes in 
my C++ software library, developed within the framework of Microsoft Visual Studio. The input comprises the original arrays (i.e., not histograms of these data).

The first test concerns the symmetry of the distributions about their corresponding median values. The results of Wilcoxon's signed-rank tests \cite{Wilcoxon1945} are given in Table \ref{tab:WilcoxonSignedRank}. 
On the basis of the two sets of p-values (depending on whether or not the so-called continuity correction is applied), there is no evidence in support of asymmetrical distributions about their corresponding 
median values.

\vspace{0.5cm}
\begin{table}[h!]
{\bf \caption{\label{tab:WilcoxonSignedRank}}}The results of Wilcoxon's signed-rank (two-tailed) tests \cite{Wilcoxon1945} on the four distributions of parental and filial heights; owing to the largeness of 
the populations, the Edgeworth approximation \cite{Kolassa1995} is reliable (and has been used). The quantity $n$ is the dimension of each input dataset. The T-domain and the T-value are standard quantities 
associated with the test. The two sets of p-values, depending on whether or not the so-called continuity correction is applied, are nearly identical.
\vspace{0.25cm}
\begin{center}
\begin{tabular}{|l|c|c|c|c|}
\hline
Quantity & Fathers & Mothers & Sons & Daughters\\
\hline
\hline
$n$ & $205$ & $205$ & $481$ & $453$\\
T-domain & $[0,21\,115]$ & $[0,21\,115]$ & $[0,115\,921]$ & $[0,102\,831]$\\
T-value & $9\,244.5$ & $8\,881.0$ & $59\,025.5$ & $45\,352.0$\\
\hline
\multicolumn{5}{|c|}{Without continuity correction}\\
\hline
p-value & $3.849 \cdot 10^{-1}$ & $7.021 \cdot 10^{-1}$ & $4.512 \cdot 10^{-1}$ & $4.180 \cdot 10^{-1}$\\
\hline
\multicolumn{5}{|c|}{With continuity correction}\\
\hline
p-value & $3.852 \cdot 10^{-1}$ & $7.027 \cdot 10^{-1}$ & $4.513 \cdot 10^{-1}$ & $4.181 \cdot 10^{-1}$\\
\hline
\end{tabular}
\end{center}
\vspace{0.5cm}
\end{table}

To test the similarity of the distributions of the height between a) fathers and sons, and b) mothers and daughters, two tests were performed: the Mann-Whitney-Wilcoxon test \cite{Wilcoxon1945,Mann1947} and 
the Brunner-Munzel test \cite{Brunner2000}. The p-values of these two tests on the data at hand are nearly identical, see Table \ref{tab:Similarity}: as a result, there is no evidence in support of dissimilar 
distributions of the height between a) fathers and sons, and b) mothers and daughters. Therefore, on the basis of his data, no evidence can be produced against Galton's first conjecture about the similarity 
of the two (one for male, another for female subjects) PDFs of the human height~\footnote{Without doubt, the mean height of the humans gradually increased during the twentieth century, e.g., see Figs.~6-8 of 
Ref.~\cite{NCD2016}. However, this effect has been associated with better nutrition and improved healthcare. According to Ref.~\cite{Roser2013}: ``Poor nutrition and illness in childhood limit human growth. 
As a consequence, the mean height of a population is strongly correlated with living standards in a population.''} ``from one generation to another'' (see beginning of Section \ref{sec:Galton}).

\vspace{0.5cm}
\begin{table}[h!]
{\bf \caption{\label{tab:Similarity}}}Results of the two (two-tailed) tests for similarity of the distributions of the height between a) fathers and sons, and b) mothers and daughters. The quantities $n$ 
and $m$ are the dimensions of the input datasets (parent and filial descendant) in each case. The U-domain, U-value, and T-value are standard quantities associated with the two tests. 
\vspace{0.25cm}
\begin{center}
\begin{tabular}{|l|c|c|}
\hline
Quantity & Fathers, sons & Mothers, daughters\\
\hline
\hline
$n$ & $205$ & $205$\\
$m$ & $481$ & $453$\\
Median height parent & $69.5^{\prime\prime}$ & $64.0^{\prime\prime}$\\
Median height offspring & $69.2^{\prime\prime}$ & $64.0^{\prime\prime}$\\
\hline
\multicolumn{3}{|c|}{Results of the Mann-Whitney-Wilcoxon test \cite{Wilcoxon1945,Mann1947}}\\
\hline
U-domain & $[0,98\,605]$ & $[0,92\,865]$\\
U-value & $49\,939.5$ & $46\,094.0$\\
p-value, without continuity correction & $7.881 \cdot 10^{-1}$ & $8.805 \cdot 10^{-1}$\\
p-value, with continuity correction & $7.883 \cdot 10^{-1}$ & $8.807 \cdot 10^{-1}$\\
\hline
\multicolumn{3}{|c|}{Results of the Brunner-Munzel test \cite{Brunner2000}}\\
\hline
T-value & $0.2702$ & $-0.1505$\\
p-value & $7.871 \cdot 10^{-1}$ & $8.805 \cdot 10^{-1}$\\
\hline
\end{tabular}
\end{center}
\vspace{0.5cm}
\end{table}

The aforementioned tests suggest that the distributions of the height a) of the fathers and of the sons, and b) of the mothers and of the daughters may be combined into two distributions: one for the male, 
another for the female subjects. To rid the data of the rounding effects~\footnote{The height of $24$ (out of $205$) fathers in Galton's family data is given as $70^{\prime\prime}$. Such shortcomings would 
hardly have occurred, if the data acquisition had been carried out by professionals.} (which give rise to artefacts in the analysis), random uniformly-distributed noise U(0,1), offset by $0.5$ so that its 
expectation value would correspond to vanishing correction, was added to all available measurements of the height; in essence, this operation neutralises Galton's concern that ``the heights are commonly 
given only to the nearest inch.'' Two new datafiles were created after the addition of the `noise', one for the male, another for the female subjects, and the resulting distributions were tested for normality.

\begin{figure}
\begin{center}
\includegraphics [width=15.5cm] {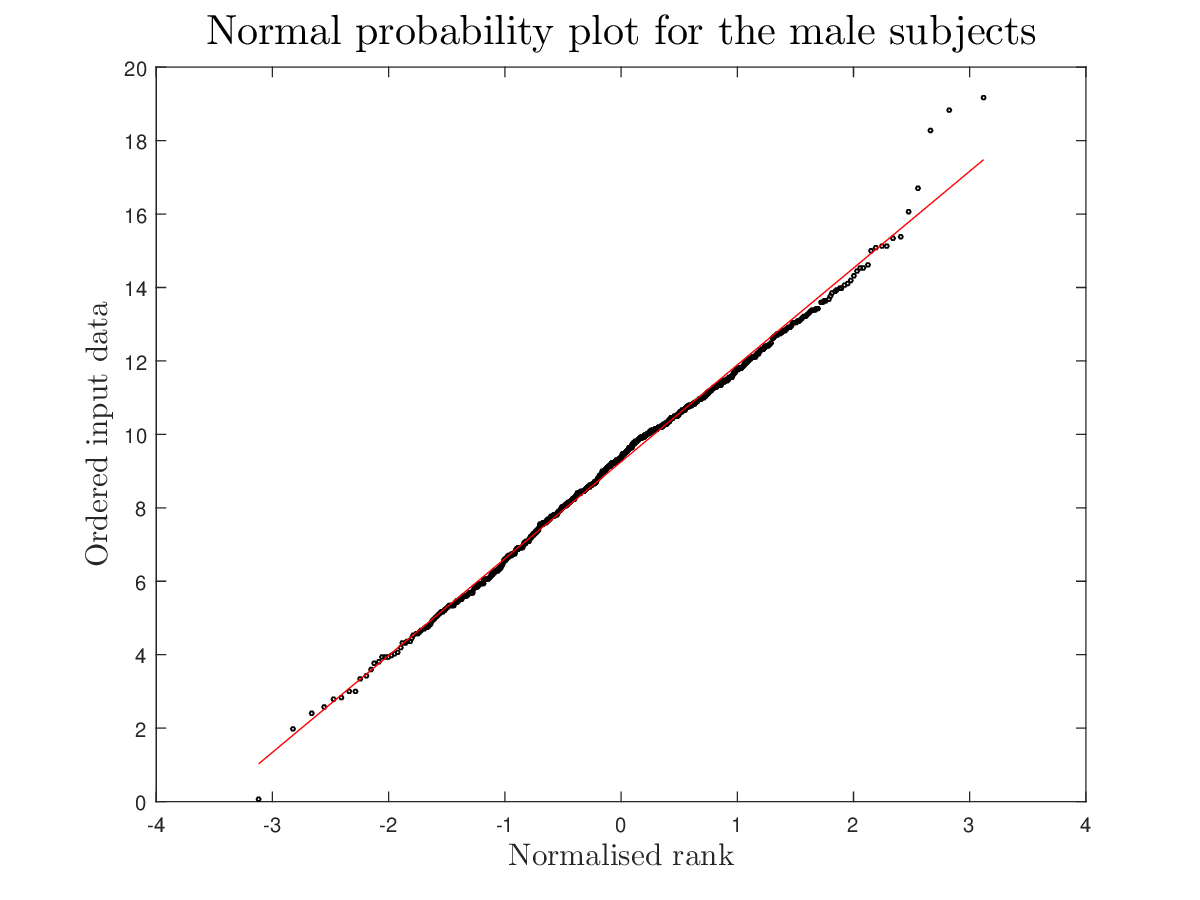}
\caption{\label{fig:NPPMaleSubjects}The normal probability plot for the male subjects (fathers and sons) in Galton's family data. The red straight line is the result of an optimisation with equal statistical 
weights for all datapoints (see Appendix \ref{App:AppA2}). The departure of the datapoints from linearity in the normal probability plot is equivalent to the departure of the input data from normality.}
\vspace{0.5cm}
\end{center}
\end{figure}

\begin{figure}
\begin{center}
\includegraphics [width=15.5cm] {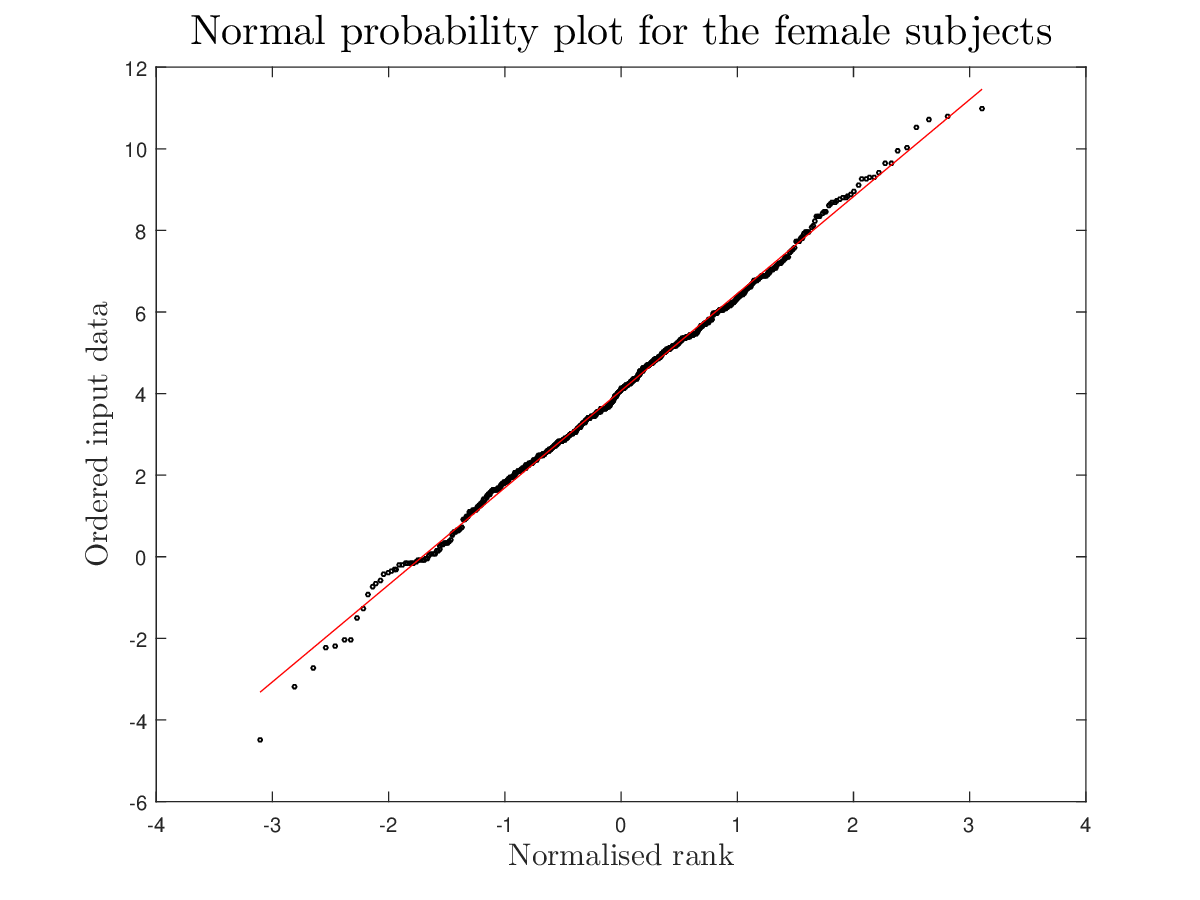}
\caption{\label{fig:NPPFemaleSubjects}The normal probability plot for the female subjects (mothers and daughters) in Galton's family data, see also caption of Fig.~\ref{fig:NPPMaleSubjects}.}
\vspace{0.5cm}
\end{center}
\end{figure}

\begin{figure}
\begin{center}
\includegraphics [width=15.5cm] {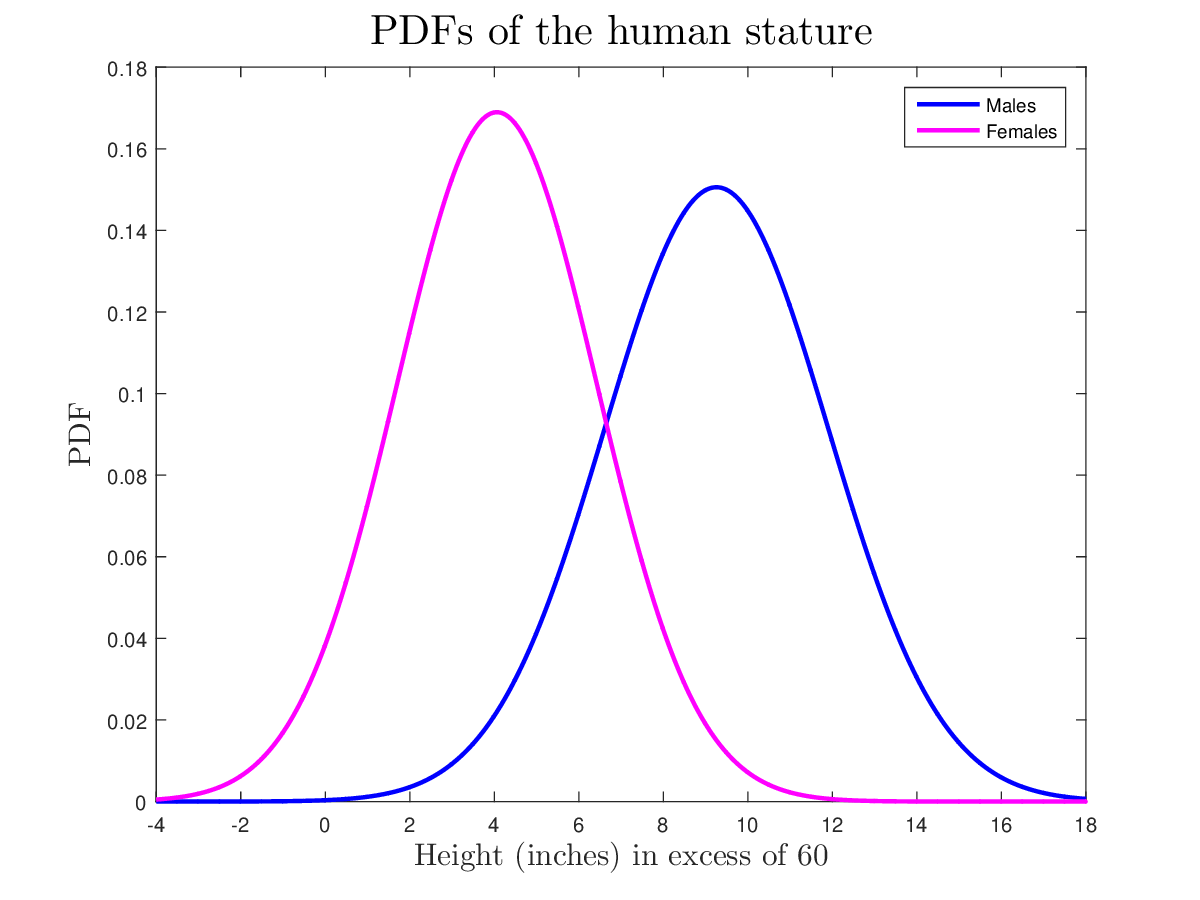}
\caption{\label{fig:HumanStature}The two PDFs of the height corresponding to the male (father and sons) and the female (mother and daughters) subjects in Galton's family data. It must be borne in mind that 
these two distributions represent ``British subjects, resident in the United Kingdom,'' who decided to send the ``best extracts from their own family records'' to Galton in 1884.}
\vspace{0.5cm}
\end{center}
\end{figure}

Numerous algorithms are available for testing the normality of distributions, e.g., see Refs.~\cite{Shapiro2015,DAgostino1990} and the works cited therein. In this study, the normality will be tested by 
means of three well-established statistical methods.
\begin{itemize}
\item The formal Shapiro-Wilk normality test, which emerges as the ultimate test for normality in power studies (e.g., see Ref.~\cite{MohdRazali2011}), was introduced in 1965 \cite{Shapiro1965} for small 
samples (in the first version of the algorithm, a maximum of $50$ observations could be tested) and was substantially extended (for application to large samples, certainly up to $n=5\,000$ observations, 
perhaps to even larger samples) in a series of studies by Royston \cite{Royston1982-1995}.
\item The Anderson-Darling test \cite{Anderson1952} with the D'Agostino 1986 addition \cite{DAgostino1986}.
\item D'Agostino's (or the D'Agostino-Pearson) $K^2$ test, which was introduced in 1973 \cite{DAgostino1973} and appeared in its current form in 1990 \cite{DAgostino1990}.
\end{itemize}
There is only one commonality between these three tests, the obvious one: the tests result in an estimate for the p-value for the acceptance of the null hypothesis, i.e., that the underlying distribution is 
the normal distribution $N(\mu,\sigma^2)$. Without ado, I present the results of the tests in Table \ref{tab:Normality}. The normal probability plots, corresponding to the male and to the female subjects, 
are given in Figs.~\ref{fig:NPPMaleSubjects} and \ref{fig:NPPFemaleSubjects}, respectively. In both cases, a few outliers may be seen in the tails of the two distributions. Although such datapoints may be 
excluded in a more thorough analysis of Galton's family data, it was decided to leave them in the database of this work. On the basis of Table \ref{tab:Normality}, as well as of Figs.~\ref{fig:NPPMaleSubjects} 
and \ref{fig:NPPFemaleSubjects}, no evidence can be produced in support of a departure of the two distributions from normality. The two PDFs of the height corresponding to Galton's male (father and sons) 
and female (mother and daughters) subjects, obtained from the means and the rms values of Table \ref{tab:Normality}, are shown (side by side) in Fig.~\ref{fig:HumanStature}.

\vspace{0.5cm}
\begin{table}[h!]
{\bf \caption{\label{tab:Normality}}}Results of the three tests for normality of the distributions of the height of the male (fathers and sons) and of the female (mothers and daughters) subjects. The 
quantity $n$ is the dimension of the input datasets (parent and filial descendant) in each case. The $W$-statistic, $A^2$-statistic, and $K^2$-statistic are the test statistics associated with the three 
tests.
\vspace{0.25cm}
\begin{center}
\begin{tabular}{|l|c|c|}
\hline
Quantity & Fathers + sons & Mothers + daughters\\
\hline
\hline
$n$ & $686$ & $658$\\
Mean value & $69.254^{\prime\prime}$ & $64.071^{\prime\prime}$\\
Standard deviation & $2.636^{\prime\prime}$ & $2.375^{\prime\prime}$\\
Skewness & $0.001$ & $-0.013$\\
Kurtosis & $3.361$ & $3.192$\\
Excess kurtosis & $0.361$ & $0.192$\\
\hline
\multicolumn{3}{|c|}{Results of the Shapiro-Wilk test \cite{Shapiro1965,Royston1982-1995}}\\
\hline
$W$-statistic & $0.9958$ & $0.9976$\\
p-value & $6.45 \cdot 10^{-2}$ & $4.73 \cdot 10^{-1}$\\
\hline
\multicolumn{3}{|c|}{Results of the Anderson-Darling test \cite{Anderson1952,DAgostino1986}}\\
\hline
$A^2$-statistic & $0.6594$ & $0.3764$\\
p-value & $8.52 \cdot 10^{-2}$ & $4.11 \cdot 10^{-1}$\\
\hline
\multicolumn{3}{|c|}{Results of the $K^2$ test \cite{DAgostino1990,DAgostino1973}}\\
\hline
$K^2$-statistic & $3.2791$ & $1.1738$\\
p-value & $1.94 \cdot 10^{-1}$ & $5.56 \cdot 10^{-1}$\\
\hline
\end{tabular}
\end{center}
\vspace{0.5cm}
\end{table}

\subsubsection{\label{sec:GaltonModelling}Modelling}

Regarding the modelling of Galton's family data, two approaches would appeal to the physicist, both involving the extraction of the mean filial heights (for the sake of simplicity, both will be denoted as 
$y_i$, but will be separately modelled) for each family (of course, families without any sons or without any daughters do not contribute to the corresponding datafile). Provided that the filial heights are 
sampled from normal distributions with two family-dependent means and two family-independent variances~\footnote{In the language of the analysis of variance (ANOVA), these variances are associated with the 
\emph{within-treatments} variations.}, i.e., one variance relevant to the heights of the sons $\sigma^2_M$, another to those of the daughters $\sigma^2_F$, the two standard errors of the means $\delta y_i$ 
are expected to be equal to $\sigma_X / \sqrt{n_i}$, where $\sigma^2_X$ is to be identified either with $\sigma^2_M$ or with $\sigma^2_F$, and $n_i$ stands for the fraternal or sororal multiplicity in the 
$i$-th family. As explained in Appendix \ref{App:AppA3}, a statistical weight $w_i$ is introduced, to account for the uncertainty $\delta y_i$ of each of the two filial means: evidently, $w_i = n_i / \sigma^2_X$, 
and, as the plan is to perform separate fits to the two types of mean filial heights, one may simply use $w_i = n_i$ for the purposes of the optimisation: the rescaling of the statistical weights (i.e., 
their multiplication by any constant $\in \mathbb{R}_{>0}$) has no impact on the important results of the fits.

I shall next address the two types of optimisation to be pursued in relation to Galton's family data. The first one involves multiple linear regression (with the parental heights as the two independent 
variables), as detailed in Appendix \ref{App:AppA5}. The second possibility involves the combination of the two parental heights into one value, i.e., the one corresponding to the hypothetical mid-parent 
in Galton's language. However, the realisation of this possibility does not involve Galton's simple procedure, but a more rigorous one, namely the separate standardisation of the two paternal heights: the 
notion of tallness can be quantified (in the absolute sense) on the basis of a comparison of each parent's height with the heights of the parents of the same sex. To this end, one may obtain an absolute 
measure of tallness for each of the two parents of the $i$-th family by introducing the quantity $z_i$ with the relation
\begin{equation} \label{eq:EQ005}
z_i = \frac{h_i - \bar{h}}{\sqrt{{\rm var} (h)}} \, \, \, ,
\end{equation}
where $h_i$ denotes the height of the parent, and $\bar{h}$ and ${\rm var} (h)$ stand for the mean height and for the (unbiased) variance of the distribution of the height of the parents of the same sex; 
although the values of Table \ref{tab:GaltonHeights} could be used, I decided to employ the statistical measures of the two distributions of parental height for those of the families with at least one son 
(when modelling the fraternal heights) and at least one daughter (when modelling the sororal heights), see Table \ref{tab:GaltonHeightsSpecific}.

\vspace{0.5cm}
\begin{table}[h!]
{\bf \caption{\label{tab:GaltonHeightsSpecific}}}Approximate mean and rms values of the distributions of parental heights corresponding to Galton's $179$ families with at least one son (suitable for the 
modelling of the fraternal heights) and the $176$ families with at least one daughter (suitable for the modelling of the sororal heights).
\vspace{0.25cm}
\begin{center}
\begin{tabular}{|l|c|c|}
\hline
Family member & Mean value & rms\\
\hline
\hline
\multicolumn{3}{|c|}{Obtained from the data of the $179$ families with at least one son}\\
\hline
Father & $69.099^{\prime\prime}$ & $2.547^{\prime\prime}$\\
Mother & $63.994^{\prime\prime}$ & $2.367^{\prime\prime}$\\
\hline
\multicolumn{3}{|c|}{Obtained from the data of the $176$ families with at least one daughter}\\
\hline
Father & $69.418^{\prime\prime}$ & $2.732^{\prime\prime}$\\
Mother & $64.125^{\prime\prime}$ & $2.289^{\prime\prime}$\\
\hline
\end{tabular}
\end{center}
\vspace{0.5cm}
\end{table}

After standardising the height of each parent of each family (using separate distributions, as explained above), one may either proceed to add the two $z$-scores or to average them (the choice in this work), 
and assign that score (independent variable) to the mid-parent for the purposes of a WLR, as detailed in Appendix \ref{App:AppA1}.

Before advancing, I should mention that I have not studied the papers relevant to the past analyses of Galton's family data; this work aims at applying a few standard methods of linear regression, not at 
compiling a review article on Galton's dataset. The interested reader is addressed to Ref.~\cite{Hanley2004}, as well as to the works cited therein.

\emph{2.2.3.1 Results of the optimisation in case of multiple linear regression}

The results of linear regression, using two independent variables (the paternal heights $x_i$ and the maternal heights $z_i$), are given in Table \ref{tab:GaltonModellingMultiple} for the two fits to the 
filial heights. The square of the Birge factor \cite{Birge1932} is an estimate for the (unbiased) `unexplained' variance in each case, see also Appendices \ref{App:AppA3}, \ref{App:AppB}, and \ref{App:AppC}.

\vspace{0.5cm}
\begin{table}[h!]
{\bf \caption{\label{tab:GaltonModellingMultiple}}}The results of linear regression, using two independent variables (the paternal heights $x_i$ and the maternal heights $z_i$) for the separate fits to the 
filial heights. All quantities have been defined in Appendix \ref{App:AppA5}. Due to the largeness of the two samples, the values of the t-multiplier are small, and modify (to the quoted precision) only the 
fitted uncertainty of $p_{1x}$ (from $0.041$ to $0.042$) in case of the fit to the sororal heights.
\vspace{0.25cm}
\begin{center}
\begin{tabular}{|l|c|c|}
\hline
Quantity & Sons & Daughters\\
\hline
\hline
$N$ & $179$ & $176$\\
$F(\hat{p}_0,\hat{p}_{1x},\hat{p}_{1z})$ & $1\,236.75$ & $941.58$\\
Birge factor & $2.65$ & $2.33$\\
$\hat{p}_0$ & $19.3^{\prime\prime}$ & $18.8^{\prime\prime}$\\
$\hat{p}_{1x}$ & $0.418$ & $0.374$\\
$\hat{p}_{1z}$ & $0.329$ & $0.304$\\
$\delta \hat{p}_0$ & $4.7^{\prime\prime}$ & $4.2^{\prime\prime}$\\
$\delta \hat{p}_{1x}$ & $0.053$ & $0.041$\\
$\delta \hat{p}_{1z}$ & $0.052$ & $0.049$\\
t-multiplier & $1.0028$ & $1.0029$\\
\hline
\end{tabular}
\end{center}
\vspace{0.5cm}
\end{table}

\begin{figure}
\begin{center}
\includegraphics [width=15.5cm] {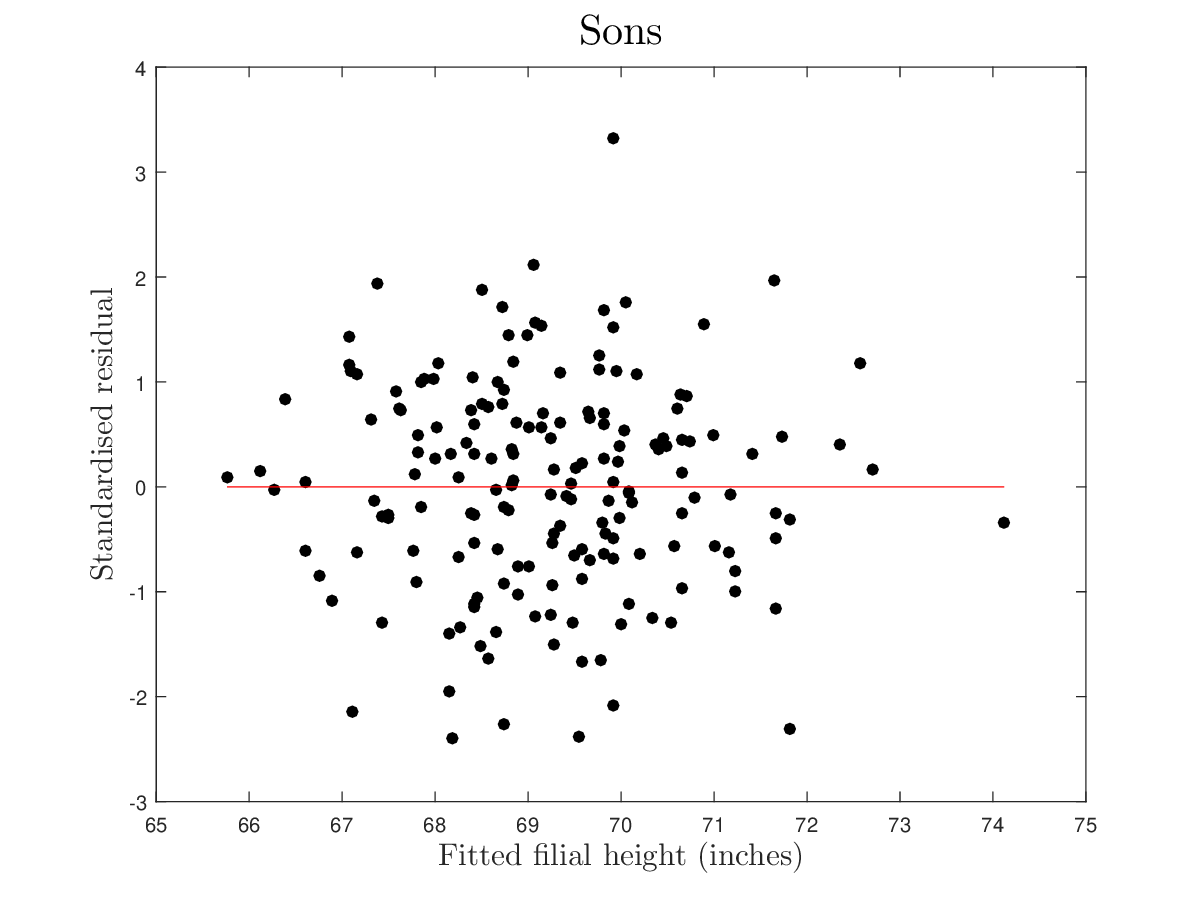}
\caption{\label{fig:GaltonMultipleRegressionSons}Scatter plot of the standardised residuals $(y_i - \tilde{y}_i) / \delta y_i$ versus the fitted values $\tilde{y}_i$ for the fit to the mean fraternal heights 
$y_i$. The red straight line corresponds to the ideal description of the input data.}
\vspace{0.5cm}
\end{center}
\end{figure}

\begin{figure}
\begin{center}
\includegraphics [width=15.5cm] {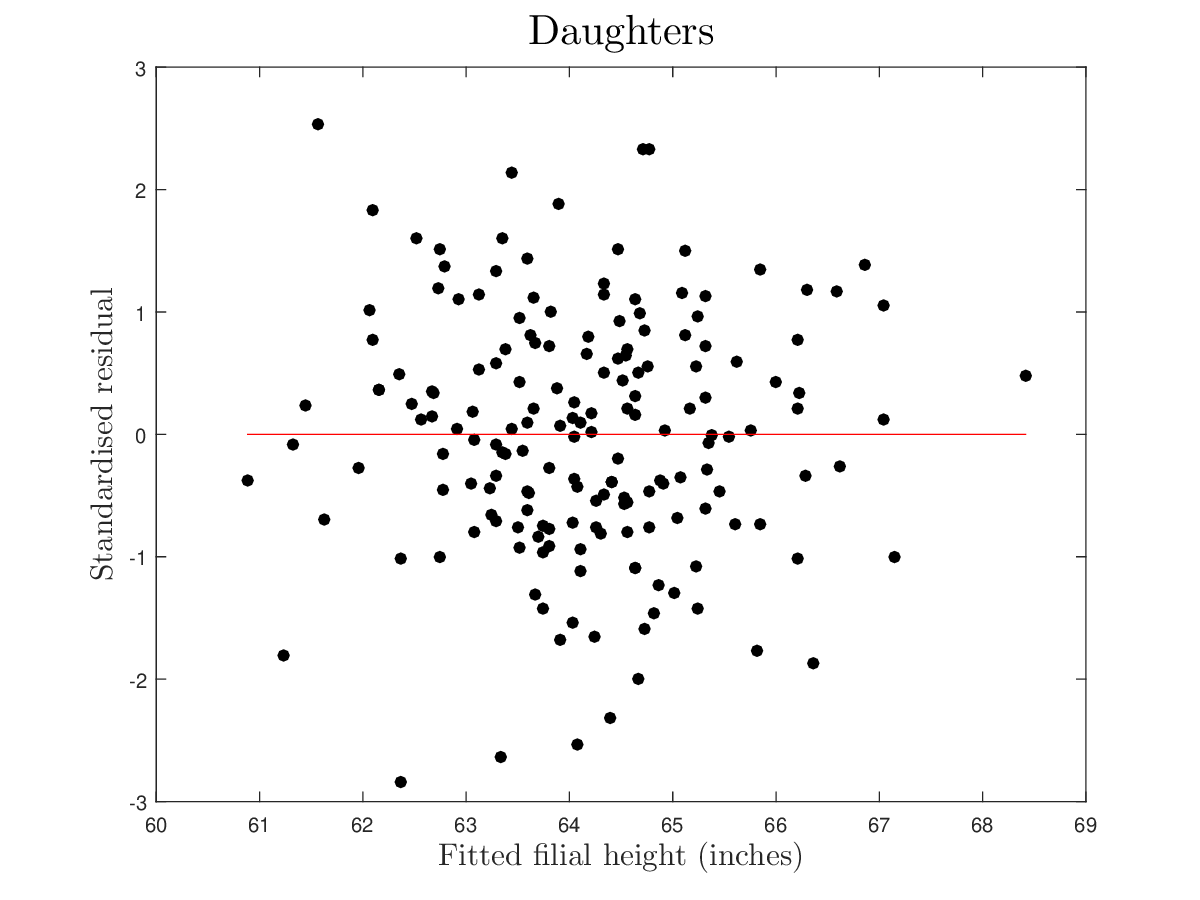}
\caption{\label{fig:GaltonMultipleRegressionDaughters}The equivalent of Fig.~\ref{fig:GaltonMultipleRegressionSons} for the fit to the mean sororal heights.}
\vspace{0.5cm}
\end{center}
\end{figure}

Scatter plots of the standardised residuals $(y_i - \tilde{y}_i) / \delta y_i$ versus the fitted values $\tilde{y}_i$ are given in Figs.~\ref{fig:GaltonMultipleRegressionSons} and 
\ref{fig:GaltonMultipleRegressionDaughters} for the fits to the fraternal and sororal heights, respectively.

\emph{2.2.3.2 Results of the optimisation in case of a WLR}

To start with, the mean filial heights correlate equally well with the mid-parents' $z$-scores: Pearson's correlation coefficient is equal to $0.593$ in case of the mean fraternal heights and $0.592$ in case 
of the mean sororal heights. The results for the fitted values of the parameters and of their uncertainties, obtained from the two fits to the available data, are given in Table \ref{tab:GaltonModellingWLR}. 
The quality of the two fits is nearly so good as that of the corresponding fits in case of the multiple linear regression of the previous section.

\vspace{0.5cm}
\begin{table}[h!]
{\bf \caption{\label{tab:GaltonModellingWLR}}}The results of a WLR, using as independent variable the average $z$-scores of the parental heights and as dependent variable the filial heights. All quantities 
have been defined in Appendix \ref{App:AppA5}. Due to the largeness of the two samples, the values of the t-multiplier are small, and do not modify (to the quoted precision) the fitted uncertainties of the 
two parameters of the fit.
\vspace{0.25cm}
\begin{center}
\begin{tabular}{|l|c|c|}
\hline
Quantity & Sons & Daughters\\
\hline
\hline
$N$ & $179$ & $176$\\
$F(\hat{p}_0,\hat{p}_1)$ & $1\,252.62$ & $963.67$\\
Birge factor & $2.66$ & $2.35$\\
$\hat{p}_0$ & $69.20^{\prime\prime}$ & $64.14^{\prime\prime}$\\
$\hat{p}_1$ & $1.82^{\prime\prime}$ & $1.71^{\prime\prime}$\\
$\delta \hat{p}_0$ & $0.12^{\prime\prime}$ & $0.11^{\prime\prime}$\\
$\delta \hat{p}_1$ & $0.17^{\prime\prime}$ & $0.16^{\prime\prime}$\\
t-multiplier & $1.0028$ & $1.0029$\\
\hline
\end{tabular}
\end{center}
\vspace{0.5cm}
\end{table}

\begin{figure}
\begin{center}
\includegraphics [width=15.5cm] {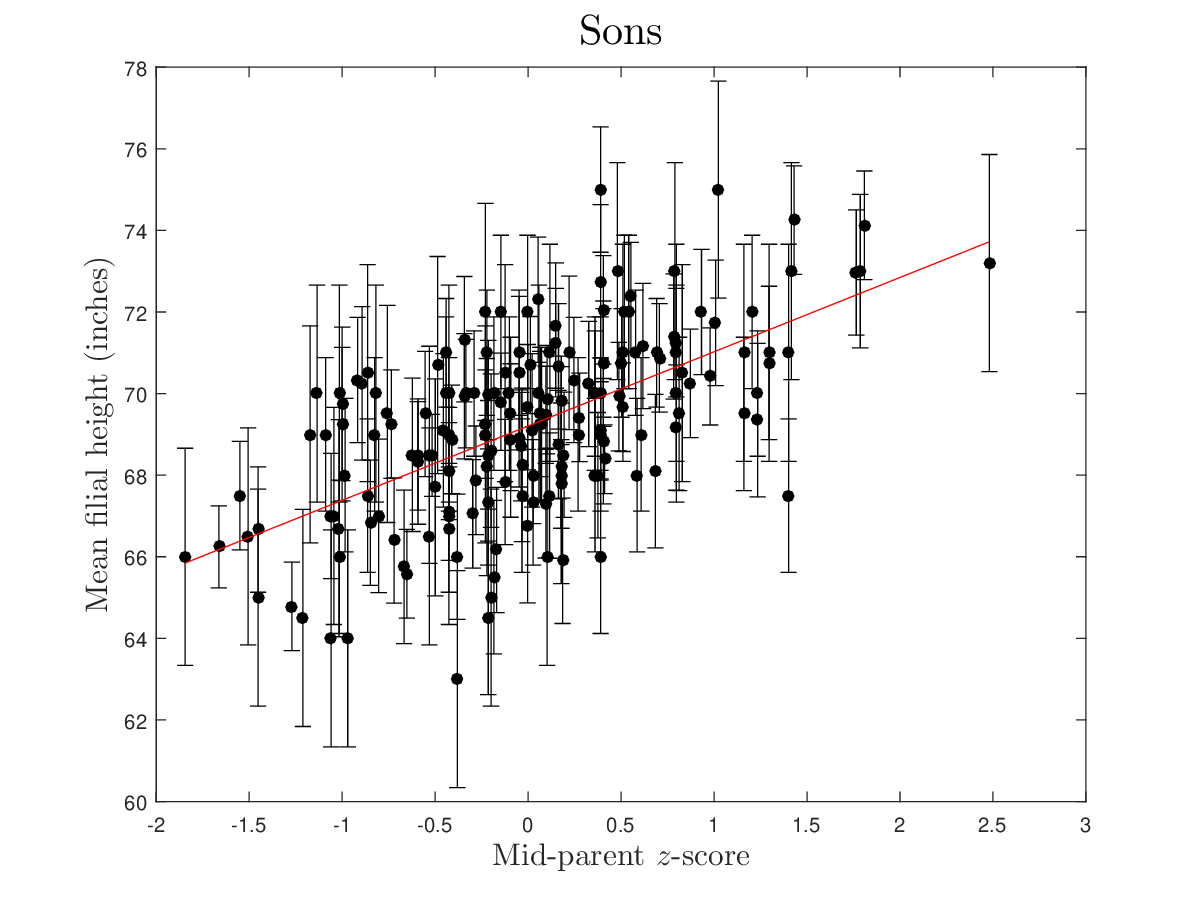}
\caption{\label{fig:GaltonWLRSons}The values of the mean fraternal height for the $179$ families with at least one son, accompanied by uncertainties corresponding to the standard error of the means. The red 
straight line represents the optimal result of the WLR, see Table \ref{tab:GaltonModellingWLR}. The abscissa is the mid-parent's $z$-score.}
\vspace{0.5cm}
\end{center}
\end{figure}

\begin{figure}
\begin{center}
\includegraphics [width=15.5cm] {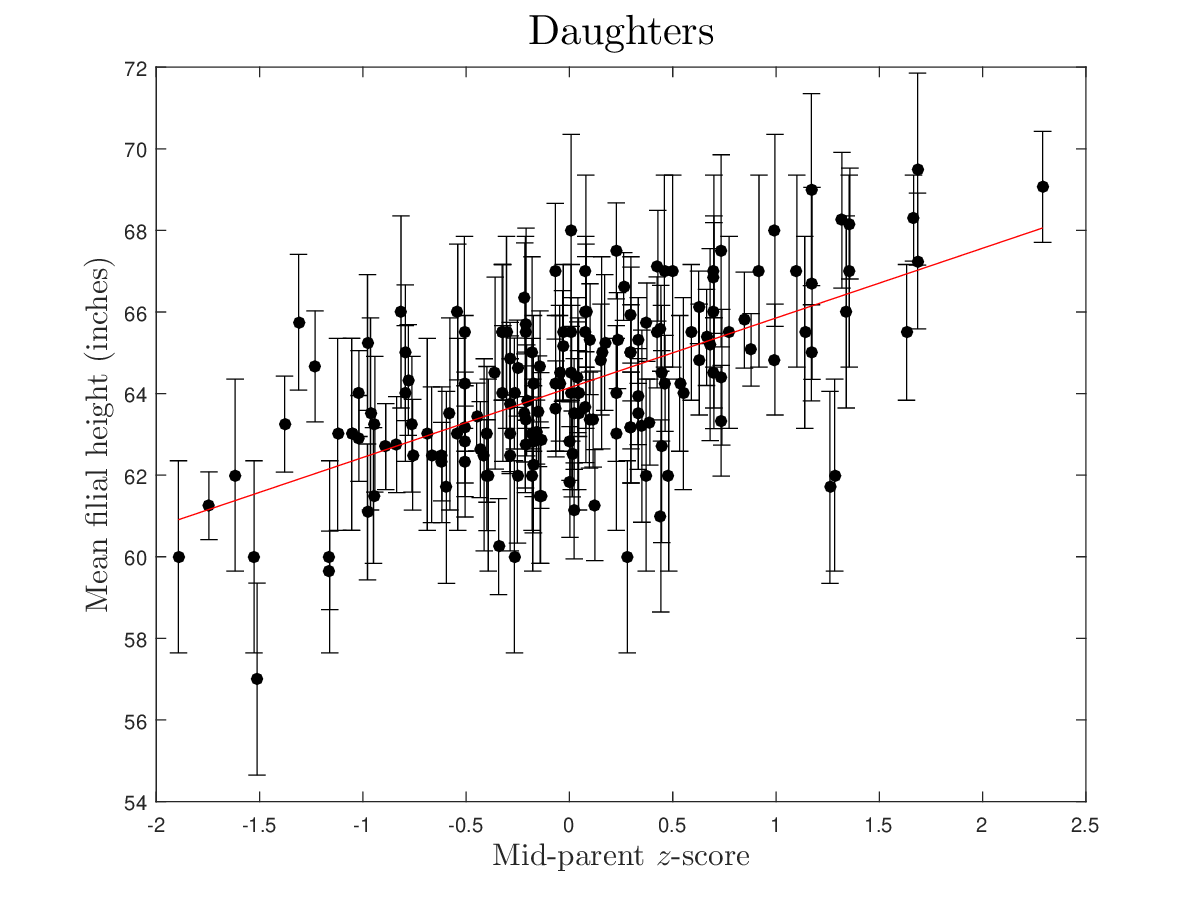}
\caption{\label{fig:GaltonWLRDaughters}The equivalent of Fig.~\ref{fig:GaltonWLRSons} for the daughters; there are $176$ families in Galton's family data with at least one daughter.}
\vspace{0.5cm}
\end{center}
\end{figure}

Plots of the mean filial heights versus the mid-parents' $z$-scores are shown in Figs.~\ref{fig:GaltonWLRSons} and \ref{fig:GaltonWLRDaughters} for the sons and the daughters, respectively.

I left one issue for the very end. Was Galton's approach to `transmuting females into male equivalents' deficient? To answer this question, let us consider two populations - one of male, another of female 
subjects - which fulfil two conditions.
\begin{itemize}
\item The dimensions of the populations are equal.
\item So far as human stature is concerned, the correspondence between the two groups is bijective. In addition, for each male subject with height $x_i$, there exists a female subject with height $z_i=x_i/k$, 
where $k$ is a constant.
\end{itemize}
In such a population, the means of the heights of the male subjects $\bar{x}$ and of the female subjects $\bar{z}$ are related: $\bar{z} = \bar{x} / k$; similarly linked are the rms values of the two 
distributions, i.e., $\sigma_x$ for the male subjects and $\sigma_z$ for the female subjects: $\sigma_z = \sigma_x / k$. Let us next focus on one family, with parental heights $x_a$ and $z_b$. Galton 
suggested that each mid-parent's height $m_{ab}$ should be taken as
\begin{equation} \label{eq:EQ006}
m_{ab} = \frac{x_a + k z_b}{2} \, \, \, .
\end{equation}
On the other hand, the average $z$-score, corresponding to the heights of these two parents - see Eq.~(\ref{eq:EQ005}), would be equal to
\begin{equation} \label{eq:EQ007}
\frac{1}{2} \left( \frac{x_a - \bar{x}}{\sigma_x} + \frac{z_b - \bar{z}}{\sigma_z} \right) \, \, \, ,
\end{equation}
which yields
\begin{equation} \label{eq:EQ008}
\frac{1}{2} \left( \frac{x_a - \bar{x}}{\sigma_x} + \frac{z_b - \bar{x} / k}{\sigma_x / k} \right) = \frac{x_a + k z_b - 2 \bar{x}}{2 \sigma_x} = \frac{m_{ab} - \bar{x}}{\sigma_x} \, \, \, .
\end{equation}

Given the constancy of the quantities $\bar{x}$ and $\sigma_x$, there can be no doubt that, provided that the two aforementioned conditions about the two populations are fulfilled, Galton's mid-parent approach 
and the method involving the average $z$-score of the parental heights, are bound to give \emph{identical} results so far as the quality of the description of the filial heights is concerned. In case of 
Galton's family data, the first criterion is undoubtedly fulfilled: the $205$ families contain $205$ fathers and $205$ mothers. However, the second criterion is not fulfilled (in the mathematical sense): 
given that the heights of the subjects are sampled from continuous probability distributions, one cannot expect to find any two subjects in the entire population, whose heights could possibly have a ratio 
\emph{exactly} equal to the ratio $k$ of the means of the heights in the two populations of male and female subjects; the consequence is that $\sigma_z \neq \sigma_x / k$, and the average $z$-score of the 
parental heights is no longer given by Eq.~(\ref{eq:EQ008}). Only one question remains: how effective (in terms of the description of the input data) was Galton's introduction of the mid-parent's height via 
Eq.~(\ref{eq:EQ006}) as the independent variable in the linear-regression problem?

The results of the two WLRs to the average filial heights, using as independent variable Galton's mid-parent's height of Eq.~(\ref{eq:EQ006}), are given in Table \ref{tab:GaltonModellingWLR-Galton}. The 
quality of the fits is comparable to that obtained with the two methods of this work. All things considered, Galton's use of the mid-parent's height as the independent variable in the problem, which he set 
out to examine, was a reasonable approximation.

\vspace{0.5cm}
\begin{table}[h!]
{\bf \caption{\label{tab:GaltonModellingWLR-Galton}}}The results of a WLR, using as independent variable Galton's mid-parent's height of Eq.~(\ref{eq:EQ006}) and as dependent variable the filial heights. The 
optimal values of $k$ per filial type have been used: $1.080$ for the fraternal heights, $1.083$ for the sororal ones. All quantities have been defined in Appendix \ref{App:AppA5}. Due to the largeness of 
the two samples, the values of the t-multiplier are small, and modify (to the quoted precision) only the fitted uncertainty of $p_{1}$ (from $0.068$ to $0.069$) in case of the fit to the fraternal heights.
\vspace{0.25cm}
\begin{center}
\begin{tabular}{|l|c|c|}
\hline
Quantity & Sons & Daughters\\
\hline
\hline
$N$ & $179$ & $176$\\
$F(\hat{p}_0,\hat{p}_1)$ & $1\,252.98$ & $953.79$\\
Birge factor & $2.66$ & $2.34$\\
$\hat{p}_0$ & $19.9^{\prime\prime}$ & $18.3^{\prime\prime}$\\
$\hat{p}_1$ & $0.713$ & $0.661$\\
$\delta \hat{p}_0$ & $4.7^{\prime\prime}$ & $4.2^{\prime\prime}$\\
$\delta \hat{p}_1$ & $0.068$ & $0.060$\\
t-multiplier & $1.0028$ & $1.0029$\\
\hline
\end{tabular}
\end{center}
\vspace{0.5cm}
\end{table}

\clearpage
\section{\label{sec:Conclusions}Conclusions}

This work aimed at providing a modicum of perspective into the analysis of two well-known datasets in terms of the methods of linear regression.

The first of these sets has been taken from Hubble's 1929 paper \cite{Hubble1929}; that work investigated the relation between the distances $r$ of nebulae (identified with galaxies later on) and their 
radial velocities $v$ (relative to the observer). Hubble assumed a linear relation between these two quantities, and determined the slope in their scatter plot; that slope, which became known as the `Hubble 
constant' $H_0$ at later times, was estimated by Hubble to about $450-600$ km/s Mpc$^{-1}$ \cite{Hubble1929,Hubble1931}. After revisiting Hubble's problem, using his selection of galaxies in conjunction with 
present-day $(r,v)$ data \cite{NED}, this work demonstrated that Hubble could have obtained an $H_0$ estimate close to the currently accepted value \cite{PDG2022} (which is several times smaller than his 
1929 and 1931 estimates), if accurate information (in particular, more accurate estimates for the distances $r$) had been available to him, see Table \ref{tab:HubbleUpdated} and Fig.~\ref{fig:HubbleUpdated}.

The second dataset of this work related to Galton's family data on human stature \cite{Galton1889}. Galton's original data, retrieved from digital photographs of eight pages of his notebook available from 
Ref.~\cite{GaltonOriginal}, contain records of parental and filial heights for each family separately. The four distributions of the height (two parental, two filial) were first tested for symmetry (about 
their respective median values), see first part of Section \ref{sec:GaltonDistributions}. Subsequently, the similarity of the distributions of the height for the male subjects (fathers and sons), as well 
as for the female subjects (mothers and daughters), was investigated and confirmed. The two distributions, obtained after combining the heights of the male subjects, as well as those of the female subjects, 
were successfully tested for normality. The two PDFs of the height, corresponding to the male and female ``British subjects, resident in the United Kingdom, who furnished Galton with the best extracts from 
their own family records'' are shown, next to one another, in Fig.~\ref{fig:HumanStature}. The filial heights were subsequently modelled in two ways:
\begin{itemize}
\item in terms of multiple linear regression, using the parental heights as independent variables and
\item in terms of a WLR, featuring the average $z$-score of the two parents as the independent variable. Each of the $z$-scores of the two parents was obtained on the basis of a comparison of that parent's 
height with the statistical measures corresponding to the distribution of the height of parents of the same sex. My thesis is that the separate standardisation of the parental heights represents the best 
option towards establishing an absolute standard of `parental tallness' in the general case.
\end{itemize}
The two methods of linear regression gave nearly identical results (in terms of the description of the input data), supporting the hypothesis of a positive correlation between parental and filial heights, 
see Tables \ref{tab:GaltonModellingMultiple} and \ref{tab:GaltonModellingWLR}, and Figs.~\ref{fig:GaltonWLRSons} and \ref{fig:GaltonWLRDaughters}.

In addition to these results, this paper may also serve as a short guide to obtaining the formalism which is relevant to the various methods of linear regression, including the determination of the optimal 
values of the parameters and of their uncertainties, as well as of any predictions based on the results of the fits: given in Appendix \ref{App:AppA} are analytical formulae for all these quantities in case 
of one independent variable, for different types of input uncertainties, see Appendices \ref{App:AppA1}-\ref{App:AppA4}. Appendix \ref{App:AppA5} deals with the simplest case of multiple linear regression, 
the one involving two independent variables. Appendices \ref{App:AppB} and \ref{App:AppC} deal with two technical issues, namely with some features of the numerical (as opposed to the analytical) 
minimisation, as well as with the application of the Birge factor to the fitted uncertainties of the parameters.

\begin{ack}
I acknowledge several interesting discussions with Davide Sardella on a number of subjects relating to Astronomy, from the very ancient to the very modern times.

Figure \ref{fig:BothUncertainties} has been created with CaRMetal \cite{CaRMetal}. The remaining figures of this paper have been created with MATLAB$\textsuperscript{\textregistered}$ (The MathWorks, Inc., 
Natick, Massachusetts, United States).
\end{ack}

\clearpage
\newpage
\appendix
\section{\label{App:AppA}Cases of linear regression relevant to this work}

The structure of this appendix is as follows. In the next section, I shall give a concise description of the WLR with one independent variable. The ordinary (called `simple' by some authors~\footnote{In my 
opinion, the adjective `simple' should rather be used in order to distinguish linear regression with one independent variable from multiple linear regression.}) linear regression with one independent 
variable will be treated in Section \ref{App:AppA2} as a special case of Section \ref{App:AppA1}. Section \ref{App:AppA3} addresses the problem of the WLR when `meaningful' (i.e., representing $1 \sigma$ 
effects in the normal distribution for each of the observations) measurement uncertainties are supplied for one variable. To deal with meaningful measurement uncertainties in both variables (a relatively 
new subject, which has not been addressed in any of the books \cite{Carroll1988,Maddala1992,Graybill1994,Hastie2016,Darlington2019}), several methods have been developed: Bayesian \cite{Gull1989}, 
errors-in-variables (e.g., see Refs.~\cite{Fuller1987,Gillard2006,Yi2021}), as well as least-squares methods \cite{Tellinghuisen2010,Tellinghuisen2020}, etc. Detailed in Section \ref{App:AppA4} are two 
relevant options, including the method which is recommended in Ref.~\cite{Tellinghuisen2020}. The simplest of multiple linear regressions, i.e., the one involving two independent variables, is discussed in 
Section \ref{App:AppA5}.

Not addressed in this appendix are established tests regarding the applicability of the aforementioned methods (e.g., whether or not linearity is established in a dataset, whether or not heteroscedasticity 
is detected in a dataset, whether or not the fitted residuals are normally distributed, and the like). The technical details of such tests are not difficult to obtain, e.g., see 
Refs.~\cite{Carroll1988,Maddala1992,Graybill1994,Hastie2016,Darlington2019}, and there is no reason for repetition in this study.

For the purposes of this appendix, the quantity $\sigma^2$ (the unbiased average of the weighted squares comprising the minimisation function $F(p_k)$, where $p_k$, $k={1, 2, \dots , N_p}$, represent the 
parameters of the fit) will stand for the ratio between the minimal value of the minimisation function $F(\hat{p}_i)$ and the NDF in the fit:
\begin{equation} \label{eq:EQA000}
\sigma^2 = \frac{F(\hat{p}_i)}{\rm NDF} \, \, \, .
\end{equation}
The quantity NDF is equal to the number of input datapoints $N$ reduced by the number of the parameters $N_p$ of the fit: ${\rm NDF}=N - N_p$. (Some authors remove one additional DoF from $N$, believing that, 
by doing so, they will obtain an unbiased variance. However, this approach is simply wrong: the variance, obtained using the relation ${\rm NDF}=N - N_p$, is already unbiased!)

\subsection{\label{App:AppA1}The WLR with one independent variable}

In a WLR, involving two quantities - namely $x$ (independent variable) and $y$ (dependent variable) - which are expected 
to be linearly related ($y = p_0 + p_1 x$), the minimisation function has the form
\begin{equation} \label{eq:EQA001}
F(p_0,p_1) = \sum_{i=1}^N w_i \left( y_i - p_0 - p_1 x_i \right)^2 \, \, \, ,
\end{equation}
where a statistical weight~\footnote{The case $w_i = 0$ is equivalent to excluding the $i$-th datapoint from the optimisation. All formulae in this work relate to datapoints which are allowed to contribute 
to the minimisation function $F(p_0,p_1)$.} $w_i \in \mathbb{R}_{>0}$, independent of the parameters $p_{0,1}$, is assigned to each pair of $N$ observations $(x_i,y_i)$. For the purposes of this section, 
the input-data structure comprises the observations $(x_i,y_i)$ and the (independent of $p_{0,1}$) weights $w_i$. If the observations $(x_i,y_i)$ correspond to measurements of physical quantities, they are - 
more often than not - accompanied by meaningful measurement uncertainties, mostly by uncertainties associated with the dependent variable ($\delta y_i$), frequently by uncertainties in both variables, and 
rarely only by uncertainties in the independent variable ($\delta x_i$); each statistical weight $w_i$ reflects the size of such measurement uncertainties.

The optimal (or fitted) values of the two parameters entering Eq.~(\ref{eq:EQA001}), namely of the intercept $p_0$ and of the slope $p_1$, correspond to the global minimum of the function $F(p_0,p_1)$, where 
two conditions are fulfilled:
\begin{equation} \label{eq:EQA002}
\frac{\partial F(p_0,p_1)}{\partial p_0} = \frac{\partial F(p_0,p_1)}{\partial p_1} = 0 \, \, \, .
\end{equation}
From now on, the values of the parameters $p_0$ and $p_1$, which correspond to the global minimum of the function $F(p_0,p_1)$, will be denoted by $\hat{p}_0$ and $\hat{p}_1$, respectively.

The application of these conditions leads to the emergence of the system of linear equations
\begin{alignat}{2} \label{eq:EQA003}
\hat{p}_0 \sum_{i=1}^N w_i &+ \hat{p}_1 \sum_{i=1}^N w_i x_i \quad && = \sum_{i=1}^N w_i y_i\nonumber\\
\hat{p}_0 \sum_{i=1}^N w_i x_i &+ \hat{p}_1 \sum_{i=1}^N w_i x_i^2 \quad && = \sum_{i=1}^N w_i x_i y_i \, \, \, ,
\end{alignat}
or, after adopting a self-explanatory notation for the sums,
\begin{alignat}{2} \label{eq:EQA004}
\hat{p}_0 S_w &+ \hat{p}_1 S_{wx} \quad && = S_{wy}\nonumber\\
\hat{p}_0 S_{wx} &+ \hat{p}_1 S_{wxx} \quad && = S_{wxy} \, \, \, .
\end{alignat}
Equations (\ref{eq:EQA004}) may be rewritten in matrix form as
\begin{equation} \label{eq:EQA005}
\left(
\begin{array}{cc}
S_w & S_{wx}\\
S_{wx} & S_{wxx}\\
\end{array}
\right)
\left(
\begin{array}{c}
\hat{p}_0\\
\hat{p}_1\\
\end{array}
\right) =
\left(
\begin{array}{c}
S_{wy}\\
S_{wxy}\\
\end{array}
\right) \, \, \, \text{or} \, \, \, {\bf G} {\bf \hat{P}} = {\bf C} \, \, \, ,
\end{equation}
where the $2 \times 2$ matrix ${\bf G}$ may be identified with half of the Hessian matrix
\begin{equation} \label{eq:EQA005a}
{\bf H_F} = \frac{\partial F(p_0,p_1)}{\partial (p_0,p_1)} \coloneqq \left(
\begin{array}{cc}
\frac{\partial^2 F(p_0,p_1)}{\partial p_0^2} & \frac{\partial^2 F(p_0,p_1)}{\partial p_0 \partial p_1}\\
\frac{\partial^2 F(p_0,p_1)}{\partial p_1 \partial p_0} & \frac{\partial^2 F(p_0,p_1)}{\partial p_1^2}\\
\end{array}
\right) \, \, \, ,
\end{equation}
while ${\bf \hat{P}}$ and ${\bf C}$ represent the column vectors of the array $\hat{p}_{0,1}$ and of the two constants on the right-hand side (rhs) of Eqs.~(\ref{eq:EQA004}), respectively. The solution of 
the matrix Eq.~(\ref{eq:EQA005}) is
\begin{equation} \label{eq:EQA006}
{\bf \hat{P}} = {\bf G}^{-1} {\bf C} \, \, \, ,
\end{equation}
where ${\bf G}^{-1}$ is evidently the inverse of the matrix ${\bf G}$. In detail, the solution reads as
\begin{equation} \label{eq:EQA007}
\left(
\begin{array}{c}
\hat{p}_0\\
\hat{p}_1\\
\end{array}
\right) = \frac{1}{\mathscr{D}}
\left(
\begin{array}{c}
S_{wy} S_{wxx} - S_{wx} S_{wxy}\\
S_w S_{wxy} - S_{wx} S_{wy}\\
\end{array}
\right) \, \, \, ,
\end{equation}
where $\mathscr{D} = {\rm det} ({\bf G}) = S_w S_{wxx} - S_{wx}^2$. A unique solution ${\bf \hat{P}}$ is obtained when $\mathscr{D} \neq 0$.

The statistical uncertainties of the two parameters of the fit may be obtained from the Taylor expansion of the function $F(p_0,p_1)$ in the vicinity of the minimum 
$F(\hat{p}_0,\hat{p}_1) = S_{wyy} - \hat{p}_0 S_{wy} - \hat{p}_1 S_{wxy}$. Given that (on account of the conditions of Eqs.~(\ref{eq:EQA002})) the first derivatives identically vanish at the minimum, one 
obtains after retaining the first two terms in the expansion:
\begin{equation} \label{eq:EQA008}
F(p_0,p_1) \approx F(\hat{p}_0,\hat{p}_1) + \frac{1}{2} \sum_{i=0}^1 \sum_{j=0}^1 (H_F)_{ij} (p_i - \hat{p}_i) (p_j - \hat{p}_j) \, \, \, ,
\end{equation}
which implies that
\begin{equation} \label{eq:EQA009}
F(p_0,p_1) \approx F(\hat{p}_0,\hat{p}_1) + \sum_{i=0}^1 \sum_{j=0}^1 G_{ij} (p_i - \hat{p}_i) (p_j - \hat{p}_j) \, \, \, .
\end{equation}
The covariance matrix of the fit takes the form
\begin{equation} \label{eq:EQA010}
{\rm cov} (p_0,p_1) = \frac{\sigma^2}{\mathscr{D}} 
\left(
\begin{array}{cc}
S_{wxx} & -S_{wx}\\
-S_{wx} & S_w\\
\end{array}
\right) \, \, \, .
\end{equation}
The statistical uncertainties of the two parameters are obtained from the diagonal elements of the covariance matrix:
\begin{equation} \label{eq:EQA011}
(\delta \hat{p}_0)^2 = \sigma^2 \frac{S_{wxx}}{\mathscr{D}}
\end{equation}
and
\begin{equation} \label{eq:EQA012}
(\delta \hat{p}_1)^2 = \sigma^2 \frac{S_w}{\mathscr{D}} \, \, \, .
\end{equation}

I shall next address two subtle issues. First, many authors associate integer multiples (e.g., $1$, $2$, $3$, and so on) of the uncertainties of Eqs.~(\ref{eq:EQA011},\ref{eq:EQA012}) with the confidence 
intervals corresponding to the same number (e.g., $1$, $2$, $3$, and so on) of $\sigma$'s in the normal distribution; the same applies to any requested new predictions $y_p$ (see next paragraph in this 
section). This approximation is poor when the results of the fit have been obtained from a small number of observations $N$. Although the aforementioned quantities $\delta \hat{p}_0$ and $\delta \hat{p}_1$ 
do represent the \emph{standard errors} of the estimators $\hat{p}_0$ and $\hat{p}_1$, the extraction of the correct confidence intervals necessitates the application of the so-called t-multiplier~\footnote{The 
quantity $\alpha$ is the (user-defined) threshold which is associated with the outset of statistical significance.} $t_{1-\alpha/2,{\rm NDF}}$, i.e., a quantity which takes account of the size of the dataset 
which yielded the results of Eqs.~(\ref{eq:EQA011},\ref{eq:EQA012}), and which follows Student's t-distribution for NDF DoFs (e.g., for $N - 2$ DoFs, if two parameters are used in the fit), see 
Refs.~\cite{Maddala1992} (pp.~78-80) and \cite{Graybill1994} (Chapter 3.6). Using the t-multiplier, the confidence intervals of the two parameters of the fit are expressed as
\begin{equation} \label{eq:EQA013}
p_k \in \, [ \, \hat{p}_k - t_{1-\alpha/2,N - 2} \, \delta \hat{p}_k \, \, , \, \, \hat{p}_k + t_{1-\alpha/2,N - 2} \, \delta \hat{p}_k \, ] \, \, \, .
\end{equation}
The importance of this correction for small samples is demonstrated in Table \ref{tab:ConfidenceIntervals}.

\vspace{0.5cm}
\begin{table}[h!]
{\bf \caption{\label{tab:ConfidenceIntervals}}}Listed in this table are a few values of the t-multiplier $t_{1-\alpha/2,N - 2}$, a quantity which must be used in the evaluation of the confidence intervals of 
the parameters $p_0$ and $p_1$, as well as in that of the confidence intervals of any requested new predictions $y_p$. This quantity depends on the significance level $\alpha$ and on the number $N$ of the 
input datapoints $(x_i,y_i)$. The corresponding values in number of $\sigma$'s in the normal distribution, which may be thought of as the limits of $t_{1-\alpha/2,N - 2}$ when $N \to \infty$, are also cited. 
This table demonstrates that, if the multiples of the standard errors of the estimators $\hat{p}_0$ and $\hat{p}_1$ represent numbers of $\sigma$'s in the normal distribution, then narrower confidence 
intervals are obtained (e.g., compare the second and last columns of this table). The confidence intervals, corresponding to the $2 \sigma$ limits in the normal distribution, are close to those favoured by 
authors in several branches of Science ($95$~\% confidence level); in Physics, $1 \sigma$ intervals are predominantly used. The values were obtained with the Excel methods NORMSDIST (CDF of the standard 
normal distribution) and TINV (Inverse CDF of Student's t-distribution).
\vspace{0.25cm}
\begin{center}
\begin{tabular}{|l|c|c|c|}
\hline
$N$ & \# of $\sigma$'s & $\alpha$ & $t_{1-\alpha/2,N - 2}$\\
\hline
\hline
$5$ & & & $1.196881$\\
$10$ & & & $1.066528$\\
$20$ & $1$ & $0.317311$ & $1.028560$\\
$40$ & & & $1.013332$\\
$80$ & & & $1.006451$\\
\hline
$5$ & & & $3.306822$\\
$10$ & & & $2.366416$\\
$20$ & $2$ & $0.045500$ & $2.148849$\\
$40$ & & & $2.067963$\\
$80$ & & & $2.032561$\\
\hline
\end{tabular}
\end{center}
\vspace{0.5cm}
\end{table}

I shall next touch upon the subject of forecasting, which, to my surprise, is not addressed in any of the books \cite{Carroll1988,Maddala1992,Graybill1994,Hastie2016,Darlington2019} in case of the WLR. Let 
us assume that the objective is to extract (from the results of the optimisation) a `prediction' for the value of the dependent variable ($y_p$) at one specific value of the independent variable ($x_p$). 
Following the equation yielding the variance of the prediction error on p.~86 of Ref.~\cite{Maddala1992}, as well as the covariance matrix of the fit of Eq.~(\ref{eq:EQA010}) of this work, one obtains the 
expectation value of the prediction $y_p$ and its standard error $\delta y_p$ as follows:
\begin{equation} \label{eq:EQA014}
y_p = \hat{p}_0 + \hat{p}_1 x_p
\end{equation}
and
\begin{equation} \label{eq:EQA015}
(\delta y_p)^2 = \sigma^2 \left( w_p^{-1} + S_w^{-1} + \frac{(x_p - \bar{x})^2}{S_{wxx} - S_w \bar{x}^2} \right) \, \, \, ,
\end{equation}
where the quantity $\bar{x}$ stands for the \emph{weighted} mean of the $x_i$ values of the input datapoints and $w_p$ is (or would be) the (expected) statistical weight of a datapoint at $x=x_p$. It must 
be mentioned that Eq.~(\ref{eq:EQA015}) yields the standard error of a \emph{new} prediction. The standard error of the \emph{fit} does not contain the first term within the brackets on the rhs of 
Eq.~(\ref{eq:EQA015}), see also Chapter 3.7 of Ref.~\cite{Maddala1992} and Eqs.~(3.6.4,3.6.5) in Chapter 3.6 of Ref.~\cite{Graybill1994} (both assuming that $w_i=1$, $\forall i$).

I shall finalise this section by giving the relevant expressions in case that the theoretical straight line contains no intercept $p_0$; datasets relevant to this part may include force-displacement 
measurements following Hooke's law, distance-velocity data of galaxies following the Lema$\hat{\rm \i}$tre-Hubble law, etc. In such a case, the minimisation function reads as
\begin{equation} \label{eq:EQA016}
F(p_1) = \sum_{i=1}^N w_i \left( y_i - p_1 x_i \right)^2 \, \, \, .
\end{equation}
The minimal value $F(\hat{p}_1) = S_{wyy} - S_{wxy}^2/S_{wxx}$, the fitted value $\hat{p}_1 = S_{wxy}/S_{wxx}$, whereas its variance
\begin{equation} \label{eq:EQA017}
(\delta \hat{p}_1)^2 = \frac{\sigma^2}{S_{wxx}} \, \, \, ,
\end{equation}
where ${\rm NDF}=N - 1$, as the linear model now contains only one parameter. Regarding the expectation value of a new prediction $y_p$ at $x=x_p$ and its standard error $\delta y_p$, Eq.~(\ref{eq:EQA014}) 
applies without the first term on the rhs (i.e., $\hat{p}_0$), whereas Eq.~(\ref{eq:EQA015}) has now the form:
\begin{equation} \label{eq:EQA017a}
(\delta y_p)^2 = \sigma^2 \left( w_p^{-1} + \frac{x_p^2}{S_{wxx}} \right) \, \, \, .
\end{equation}

\subsection{\label{App:AppA2}The linear regression with constant (or `without') weights}

If all statistical weights $w_i$ are equal (to a constant $w \neq 0$) in Section \ref{App:AppA1}, a case which is usually referred to in the standard literature in Statistics as (fulfilling) `homoscedasticity', 
or if no weights are supplied (or are relevant) in a problem, all expressions of Section \ref{App:AppA1} apply after simply replacing all $w_i$'s by $1$. It so happens that there is no dependence of any of 
the important quantities of the fit (i.e., of the optimal values of the parameters of the fit, of the standard errors of these parameters, as well as of any predictions and of their standard errors) on the 
value of the constant $w$. Only the minimisation function $F(p_0,p_1)$ is $w$-dependent, but (as the important results of the optimisation - which also involve $F(\hat{p}_0,\hat{p}_1)$ - are ratios of 
quantities containing the same powers of $w$ in the nominators and denominators) the $w$-dependence is eliminated from the useful output. This special case of linear regression, which is generally known as 
`ordinary linear regression' (OLR for short) or `ordinary linear least-squares optimisation', provides (in the eyes of many) better insight into the nature of the various quantities. As a result, most 
contributions in the standard literature start the description of the methods of linear regression by treating the OLR case (and some do not even venture any further!).

I shall next cover some of the main features of the OLR, starting from the fitted value of the slope. The determinant $\mathscr{D}$ is equal to $N S_{xx} - S_{x}^2$, which turns out to be simply 
the variance of the input $x_i$ values multiplied by $N^2$: $\mathscr{D} = N^2 {\rm var} (x)$. From Eq.~(\ref{eq:EQA007}), one obtains for the fitted value of the slope:
\begin{equation} \label{eq:EQA018}
\hat{p}_1 = \frac{N \sum_{i=1}^N x_i y_i - \left( \sum_{i=1}^N x_i \right) \left( \sum_{i=1}^N y_i \right)}{N^2 {\rm var} (x)} = \frac{{\rm cov} (x,y)}{{\rm var} (x)} \, \, \, .
\end{equation}

Pearson's correlation coefficient $\rho_{x,y} \in [-1,+1]$ between two samples (of equal dimension) corresponding to two quantities $x$ and $y$, defined by the formula
\begin{equation} \label{eq:EQA019}
\rho_{x,y} \coloneqq \frac{{\rm cov} (x,y)}{\sqrt{{\rm var} (x) \, {\rm var} (y)}} \, \, \, ,
\end{equation}
is a measure of the linear correlation between these quantities. Using the first of Eqs.~(\ref{eq:EQA003}) with all statistical weights $w_i$ set to $1$, one obtains
\begin{equation} \label{eq:EQA020}
N \hat{p}_0 + \hat{p}_1 \sum_{i=1}^N x_i = \sum_{i=1}^N y_i \Rightarrow \hat{p}_0 + \hat{p}_1 \bar{x} = \bar{y} \, \, \, .
\end{equation}
Using this equation in order to replace $\hat{p}_0$, one obtains for the minimal value of the minimisation function:
\begin{align} \label{eq:EQA021}
F(\hat{p}_0,\hat{p}_1) &= \sum_{i=1}^N \left( y_i - \hat{p}_0 - \hat{p}_1 x_i \right)^2\nonumber\\
 &= \sum_{i=1}^N \left( y_i - \bar{y} - \hat{p}_1 \left( x_i - \bar{x} \right) \right)^2\nonumber\\
 &= \sum_{i=1}^N \left( \left( y_i - \bar{y} \right)^2 + \hat{p}_1^2 \left( x_i - \bar{x} \right)^2 - 2 \hat{p}_1 \left( x_i - \bar{x} \right) \left( x_i - \bar{y} \right) \right)\nonumber\\
 &= N {\rm var} (y) + N \hat{p}_1^2 {\rm var} (x) - 2 N \hat{p}_1 {\rm cov} (x,y) \, \, \, .
\end{align}
Making use of Eqs.~(\ref{eq:EQA018},\ref{eq:EQA019}), one finally obtains
\begin{equation} \label{eq:EQA022}
\sigma^2 = {\rm var} (y) \left( 1 - \rho_{x,y}^2 \right) \, \, \, .
\end{equation}
The quantity on the left-hand side of this equation is routinely interpreted as the part of the variance of the dependent variable which remains `unexplained' (after the optimisation), whereas the fraction 
of the variance which is associated with the $\rho_{x,y}^2$ term is considered `explained', in the sense that this fraction of the total variance of the dependent variable is understood as essentially 
originating from the variation of the independent variable. Of course, all these expressions are analogous to those obtained in the WLR after replacing the numerical measures `mean', `variance', and 
`correlation' with their weighted counterparts.

\subsection{\label{App:AppA3}The WLR with meaningful measurement uncertainties only in one variable}

At least so far as physical measurements are concerned, this is the most frequent, and hence useful case. In Physics, all measurements of physical quantities are expected to be accompanied by meaningful 
uncertainties: measurements of physical quantities without uncertainties might provide a general impression about the order of magnitude of effects, but are hardly of any use in statistical analyses. As a 
result, particular attention is paid in extracting such uncertainties for the physical quantities which are under experimental exploration, i.e., uncertainties which are indicative of the likelihood of the 
departure of the measured from the true values. Unlike most other branches of Science, Physics favours $1 \sigma$ uncertainties in the normal distribution, representing a confidence level of about $68.27$~\%.

In short, the pairs of $N$ observations $(x_i,y_i)$ are replaced in most Physics studies by triplets of $N$ observations $(x_i,y_i,\delta y_i)$, and the optimisation problem reduces to a WLR with statistical 
weights $w_i \coloneqq (\delta y_i)^{-2}$. Had it not been for one additional consideration, the treatment of this subcategory would have been unremarkable. So, what makes this case special?

The fact that the measurement uncertainties $\delta y_i$ represent $1 \sigma$ effects in the normal distribution implies that each standardised residual
\begin{equation} \label{eq:EQA023}
\frac{y_i - \tilde{y}_i}{\delta y_i}
\end{equation}
is expected to follow the standard normal distribution; in the expression above, the quantity $\tilde{y}_i$ represents the fitted value at $x=x_i$, obtained via the modelling as the case might be, in 
particular, via the linear model $y = p_0 + p_1 x$ in this work. As each of the quantities under the sum in Eq.~(\ref{eq:EQA001}) follows the standard normal distribution, $F(p_0,p_1)$ is expected to follow 
the $\chi^2$ distribution with $N$ DoFs. Given that two parameters are used in the general problem of linear regression, the minimal value $F(\hat{p}_0,\hat{p}_1)$ is, in the context of this section, 
$\chi^2$-distributed with $N - 2$ DoFs.

One of the obvious advantages of the use of meaningful uncertainties is that the quality of the fit (or, equivalently, the adequacy of the linear model to account for the input dataset) may be assessed on 
the basis of the p-value~\footnote{The p-value represents the upper tail of the CDF of the relevant distribution; of the $\chi^2$ distribution in this case.} corresponding to the $F(\hat{p}_0,\hat{p}_1)$ 
result for the given DoFs.

When dealing with the subject of this section, there is one notable difference to the formalism developed earlier in order to treat the general-case WLR: it relates to the standard errors of the parameters 
of the fit, $\delta \hat{p}_0$ and $\delta \hat{p}_1$. Equations (\ref{eq:EQA011}) and (\ref{eq:EQA012}) may be rewritten as
\begin{equation} \label{eq:EQA024}
\delta \hat{p}_0 = {\rm BF} \sqrt{\frac{S_{wxx}}{\mathscr{D}}}
\end{equation}
and
\begin{equation} \label{eq:EQA025}
\delta \hat{p}_1 = {\rm BF} \sqrt{\frac{S_w}{\mathscr{D}}} \, \, \, ,
\end{equation}
where
\begin{equation} \label{eq:EQA026}
{\rm BF} = \sqrt{\sigma^2} \, \, \, .
\end{equation}

The importance of the quantity BF of Eq.~(\ref{eq:EQA026}) in the problem of linear regression was (to the best of my knowledge) first addressed by Birge in 1932 \cite{Birge1932}. Although BF should (in my 
opinion) be called `Birge factor', the Particle Data Group (PDG) use instead the plain term `scale factor' \cite{PDG2022} in their compilations of physical constants, and recommend its application (in the 
context of this section) only when ${\rm BF} > 1$; if ${\rm BF} < 1$, the factor is omitted from Eqs.~(\ref{eq:EQA024},\ref{eq:EQA025}), the intention evidently being to prevent the decrease of the 
statistical uncertainties $\delta \hat{p}_0$ and $\delta \hat{p}_1$ when the input uncertainties $\delta y_i$ have been overly generous (which may be considered a `rare event' in Physics, yet this is another 
story). Being a particle physicist, I abide by (and have nothing against) the PDG recommendation, see also Appendix \ref{App:AppC}.

Last but not least, if only meaningful uncertainties $\delta x_i$ are supplied in a problem, one may swap the roles of the quantities $x$ and $y$, and perform a WLR using the linear model: $x = q_0 + q_1 y$. 
The optimal values of the two parameters of the originally intended straight line $y = p_0 + p_1 x$ could then be retrieved from the quantities $\hat{q}_{0,1}$ as follows: 
$(\hat{p}_0, \hat{p}_1)=(- \hat{q}_0 \hat{q}_1^{-1}, \hat{q}_1^{-1})$.

\subsection{\label{App:AppA4}The WLR with meaningful measurement uncertainties in both variables}

My first effort towards obtaining a solution, when I first encountered such a problem in the mid-1980s, rested upon the use of the WLR method with statistical weights inversely proportional to the square of 
the product of the two uncertainties for each input datapoint. In that implementation, the statistical weight $w_i$ was defined by the formula:
\begin{equation} \label{eq:EQA027}
w_i \coloneqq \frac{1}{(\delta x_i \delta y_i)^2} \frac{\sum_{i=1}^N \frac{1}{(\delta y_i)^2}}{\sum_{i=1}^N \frac{1}{(\delta x_i \delta y_i)^2}} \, \, \, .
\end{equation}
Given that the statistical weights of this equation fulfil $\sum_{i=1}^N w_i \equiv \sum_{i=1}^N (\delta y_i)^{-2}$, one arrives at a situation reminiscent of Appendix \ref{App:AppA3} with redefined statistical 
weights, and could trick oneself into considering the minimal value $F(\hat{p}_0,\hat{p}_1)$ $\chi^2$-distributed with $N - 2$ DoFs. Although the statistical weights $w_i$ sum up (by construction) to the same 
constant in the two cases, those given in Eq.~(\ref{eq:EQA027}) also take account of the uncertainties $\delta x_i$. Of course, one disadvantage of using the weights of Eq.~(\ref{eq:EQA027}) is that an input 
datapoint needs to be excluded if \emph{either} of the two input uncertainties vanishes, whereas such an exclusion is necessary only if \emph{both} input uncertainties vanish in case of the following two 
methods of this section.

Several methodologies were developed during the last decades for dealing with this problem in a rigorous manner~\footnote{In particular, errors-in-variables models emerged with the principal scope of providing 
reliable solutions to this particular situation \cite{Fuller1987,Gillard2006,Yi2021}.}. A recent article \cite{Tellinghuisen2020} lists and compares the most important of these methods. Following the results 
of his analysis, the author recommends the use of the so-called `Modified effective-variance method' (EV$_2$), in which \emph{parameter-dependent} statistical weights are assigned to the input datapoints: in 
the general case, each such weight is given by
\begin{equation} \label{eq:EQA028}
w_i (p_k) \coloneqq \left( (\delta y_i)^2 + \left( \left.\frac{d y}{d x} \right| _{x = x_i} \right)^2 (\delta x_i)^2 \right)^{-1} \, \, \, ,
\end{equation}
which, in case of linear regression, takes the form
\begin{equation} \label{eq:EQA029}
w_i (p_0,p_1) \to w_i (p_1) \coloneqq \left( (\delta y_i)^2 + p_1^2 (\delta x_i)^2 \right)^{-1} \, \, \, .
\end{equation}
Owing to its obvious dependence on one of the parameters of the WLR, the application of a statistical weight of this form introduces non-linearity into a linear problem and calls for numerical minimisation, 
see Appendix \ref{App:AppB}. I shall next elaborate on the emergence of the statistical weight of Eq.~(\ref{eq:EQA029}), by allowing myself to be guided by physical intuition.

\begin{figure}
\begin{center}
\includegraphics [width=15.5cm] {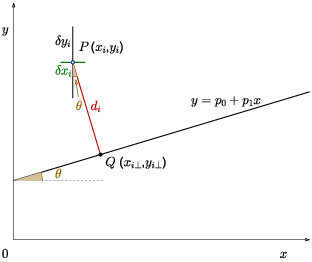}
\caption{\label{fig:BothUncertainties}The figure facilitates the development of a solution to the problem of linear regression when meaningful measurement uncertainties in both $x$ and $y$ are supplied. The 
minimisation function uses the distance $d_i$ of an input datapoint $(x_i,y_i)$ to the `current' (i.e., corresponding to a specific iteration of the numerical minimisation) straight line. The projected 
components of the two uncertainties on the orthogonal (to that straight line) direction may be quadratically (EV$_2$ method \cite{Tellinghuisen2020}) or linearly summed (see text).}
\vspace{0.5cm}
\end{center}
\end{figure}

Referring to Fig.~\ref{fig:BothUncertainties}, one may argue that the contribution of each input datapoint to the minimisation function $F(p_0,p_1)$ must involve two quantities:
\begin{itemize}
\item the minimal distance $d_i$ of that datapoint to the `current' (i.e., relating to a specific iteration of the numerical minimisation) straight line and
\item the input uncertainties $\delta x_i$ and $\delta y_i$. 
\end{itemize}
One may argue that the aforementioned contribution depends on the relative largeness of the quantity $d_i$, judged in terms of a representative combined size of the two uncertainties $\delta x_i$ and 
$\delta y_i$. To assess this, one first obtains the coordinates of point $Q$, representing the intersection of the `current' straight line and its orthogonal straight line passing through point $P$:
\begin{equation} \label{eq:EQA030}
(x_{i\perp},y_{i\perp})=\left( \frac{p_1 \left( y_i - p_0 \right) + x_i}{1+p_1^2}, \frac{p_1^2 y_i + p_1 x_i + p_0}{1+p_1^2} \right)
\end{equation}
and obtains the distance $d_i$ as follows:
\begin{equation} \label{eq:EQA031}
d_i = \frac{\left| y_i - p_0 - p_1 x_i \right|}{\sqrt{1+p_1^2}} \, \, \, .
\end{equation}
The two uncertainties $\delta x_i$ and $\delta y_i$ may be interpreted as representing different effects: $\delta y_i$ is directly associated with the statistical effects regarding the specific observation, 
whereas $\delta x_i$ may be taken to represent systematic effects, i.e., effects which are induced on $y$ as the result of the variation of the independent variable. Projected on the orthogonal direction, 
the systematic component of the uncertainty is equal to
\begin{equation} \label{eq:EQA032}
\delta x_i \left| \cos \left( \frac{\pi}{2} - \theta \right) \right| = \delta x_i \frac{\left| p_1 \right|}{\sqrt{1+p_1^2}} \, \, \, ,
\end{equation}
whereas the statistical component of the uncertainty yields
\begin{equation} \label{eq:EQA033}
\delta y_i \cos \theta = \delta y_i \frac{1}{\sqrt{1+p_1^2}} \, \, \, ;
\end{equation}
of course, $p_1 = \tan \theta$. As these two uncertainties are independent, the strict application of Gauss' law of error propagation suggests the use of the combined uncertainty in the form
\begin{equation} \label{eq:EQA034}
\sqrt{\frac{(\delta y_i)^2 + p_1^2 (\delta x_i^2)}{1+p_1^2}} \, \, \, .
\end{equation}
Therefore, the application of Gauss' law suggests the summation of the two uncertainties of Eqs.~(\ref{eq:EQA032},\ref{eq:EQA033}) in quadrature, leading to the minimisation function
\begin{equation} \label{eq:EQA034}
F(p_0,p_1) = \sum_{i=1}^N \frac{(y_i - p_0 - p_1 x_i)^2}{(\delta y_i)^2+p_1^2 (\delta x_i)^2} \equiv \sum_{i=1}^N w_i(p_1) (y_i - p_0 - p_1 x_i)^2 \, \, \, ,
\end{equation}
with $w_i(p_1)$ given by Eq.~(\ref{eq:EQA029}).

To cope with underestimated uncertainties (which, unfortunately, cannot be considered a sporadic phenomenon in Physics), many physicists favour the \emph{linear} summation of the uncertainties corresponding 
to statistical and systematic effects. For the sake of example (taken from Particle Physics), two experimental groups measured (over two decades ago) the strong-interaction shift $\epsilon_{1 s}$ 
\cite{Schroeder2001,Hennebach2014} and the total decay width $\Gamma_{1s}$ \cite{Schroeder2001,Hirtl2021} of the ground state in pionic hydrogen, utilising the same low-energy pion beamline ($\pi E 5$) at 
the Paul Scherrer Institute (PSI), the same degrading device (cyclotron trap), similar crystal-spectrometer systems, and measurements corresponding (predominantly) to the same X-ray transition ($3 p \to 1 s$). 
In spite of the obvious similarities, the earlier collaboration (the ETHZ-Neuch{\^a}tel-PSI Collaboration) recommended the linear summation of their statistical and systematic uncertainties, whereas the later 
one (the Pionic Hydrogen Collaboration) favoured the quadratic summation of their corresponding uncertainties. In essence, there is no conflict between these two choices: some researchers choose to `err on 
the side of caution' (use of the $L_1$ norm when combining statistical and systematic effects), whereas others assume a somewhat bolder stance (use of the $L_2$ norm).

In the context of this section, the linear summation of the uncertainties corresponding to statistical and systematic effects yields:
\begin{equation} \label{eq:EQA035}
w_i (p_1) = \left( \delta y_i + \left| p_1 \right| \delta x_i \right)^{-2} \, \, \, .
\end{equation}

Last but not least, if the uncertainties $\delta x_i$ are (on average) considerably larger than $\delta y_i$, it might make sense to swap the role of the two variables, see also ending paragraph of Appendix 
\ref{App:AppA3}. (Regarding Table \ref{tab:HubbleUpdated}, the MINUIT-based application yielded identical results with the two options.)

\subsection{\label{App:AppA5}The WLR with two independent variables}

Numerous articles in the literature have treated the problem of multiple linear regression. In this section, I shall detail the solution to the problem involving two independent variables $x$ and $z$, which 
is the simplest case in multiple linear regression. Assuming the linear relation $y = p_0 + p_{1x} x + p_{1z} z$, the minimisation function reads as
\begin{equation} \label{eq:EQA036}
F(p_0,p_{1x},p_{1z}) = \sum_{i=1}^N w_i \left( y_i - p_0 - p_{1x} x_i - p_{1z} z_i \right)^2 \, \, \, .
\end{equation}

The observation set now comprises $N$ quadruplets $(x_i,z_i,y_i,w_i)$, reducing to $N$ triplets $(x_i,z_i,y_i)$ for constant statistical weights $w_i = w \neq 0$, in which case the WLR with two independent 
variables becomes ordinary. Using the notation of Appendix \ref{App:AppA1}, one obtains
\begin{equation} \label{eq:EQA037}
{\bf \hat{P}} = {\bf G}^{-1} {\bf C} \, \, \, ,
\end{equation}
where ${\bf G}^{-1}$ is evidently the inverse of the symmetrical matrix
\begin{equation} \label{eq:EQA038}
{\bf G} = \left(
\begin{array}{ccc}
S_w & S_{wx} & S_{wz}\\
S_{wx} & S_{wxx} & S_{wxz}\\
S_{wz} & S_{wxz} & S_{wzz}\\
\end{array}
\right) \, \, \, ,
\end{equation}
and the column arrays ${\bf \hat{P}}$ and ${\bf C}$ are given by
\begin{equation} \label{eq:EQA039}
{\bf \hat{P}} = \left(
\begin{array}{c}
\hat{p}_0\\
\hat{p}_{1x}\\
\hat{p}_{1z}\\
\end{array}
\right)
\end{equation}
and
\begin{equation} \label{eq:EQA040}
{\bf C} = \left(
\begin{array}{c}
S_{wy}\\
S_{wxy}\\
S_{wzy}\\
\end{array}
\right) \, \, \, .
\end{equation}

After a few algebraical operations, one may obtain the inverse of the matrix ${\bf G}$ as
\begin{equation} \label{eq:EQA041}
{\bf G}^{-1} = \frac{1}{\mathscr{D}} \left(
\begin{array}{ccc}
S_{wxx} S_{wzz} - S_{wxz}^2 & \, S_{wz} S_{wxz} - S_{wx} S_{wzz} & \, S_{wx} S_{wxz} - S_{wz} S_{wxx}\\
S_{wz} S_{wxz} - S_{wx} S_{wzz} & \, S_w S_{wzz} - S_{wz}^2 & \, S_{wx} S_{wz} - S_w S_{wxz}\\
S_{wx} S_{wxz} - S_{wz} S_{wxx} & \, S_{wx} S_{wz} - S_w S_{wxz} & \, S_w S_{wxx} - S_{wx}^2\\
\end{array}
\right) \, \, \, ,
\end{equation}
where $\mathscr{D} = {\rm det} ({\bf G}) = S_w S_{wxx} S_{wzz} + 2 S_{wx} S_{wz} S_{wxz} - S_w S_{wxz}^2 - S_{wx}^2 S_{wzz} - S_{wz}^2 S_{wxx}$.

The diagonal elements of the matrix ${\bf G}^{-1}$ yield the standard errors of the parameters of the fit as follows:
\begin{align} \label{eq:EQA042}
(\delta \hat{p}_0)^2 &= \sigma^2 \frac{S_{wxx} S_{wzz} - S_{wxz}^2}{\mathscr{D}}\nonumber\\
(\delta \hat{p}_{1x})^2 &= \sigma^2 \frac{S_w S_{wzz} - S_{wz}^2}{\mathscr{D}}\nonumber\\
(\delta \hat{p}_{1z})^2 &= \sigma^2 \frac{S_w S_{wxx} - S_{wx}^2}{\mathscr{D}} \, \, \, ,
\end{align}
where $\sigma^2$ has been obtained after using ${\rm NDF}=N - 3$, as the general linear model now contains three parameters.

Following again the procedure set forth in Chapter 3.7 of Ref.~\cite{Maddala1992}, one obtains (after some effort) the expectation value of the prediction $y_p$ and its standard error $\delta y_p$ 
corresponding to $x=x_p$ and $z=z_p$ as follows:
\begin{equation} \label{eq:EQA043}
y_p = \hat{p}_0 + \hat{p}_{1x} x_p + \hat{p}_{1z} z_p
\end{equation}
and
\begin{equation} \label{eq:EQA044}
(\delta y_p)^2 = \sigma^2 \left( w_p^{-1} + S_w^{-1} + \frac{ \sigma_z^2 \left( x_p - \bar{x} \right)^2 + \sigma_x^2 \left( z_p - \bar{z} \right)^2 - 2 \rho_{x,z} \sigma_x \sigma_z \left( x_p - \bar{x} \right) \left( z_p - \bar{z} \right)}{ S_w \left( 1 - \rho_{x,z}^2 \right) \sigma_x^2 \sigma_z^2 } \right) \, \, \, ,
\end{equation}
where $\bar{x} = S_{wx} / S_w$, $\bar{z} = S_{wz} / S_w$, $\sigma_x^2 = S_{wxx} / S_w - \bar{x}^2$, and $\sigma_z^2 = S_{wzz} / S_w - \bar{z}^2$; the statistical weight $w_p$ has been explained in Section 
\ref{App:AppA1}. Finally, the weighted version of Pearson's correlation coefficient $\rho_{x,z}$ reads as
\begin{equation} \label{eq:EQA045}
\rho_{x,z} = \frac{S_{wxz} - S_w \bar{x} \bar{z}}{S_w \sigma_x \sigma_z} \, \, \, .
\end{equation}

For the sake of completeness, I shall finalise this section by giving the relevant expressions in case that the theoretical straight line is expected to contain no intercept $p_0$. In such a case, the 
minimisation function reads as
\begin{equation} \label{eq:EQA046}
F(p_{1x},p_{1z}) = \sum_{i=1}^N w_i \left( y_i - p_{1x} x_i - p_{1z} z_i \right)^2 \, \, \, .
\end{equation}
The minimal value $F(\hat{p}_{1x},\hat{p}_{1z}) = S_{wyy} - \hat{p}_{1x} S_{wxy} - \hat{p}_{1z} S_{wzy}$, whereas the fitted values of the parameters $p_{1x}$ and $p_{1z}$ are as follows:
\begin{equation} \label{eq:EQA047}
{\bf \hat{P}} = \left(
\begin{array}{c}
\hat{p}_{1x}\\
\hat{p}_{1z}\\
\end{array}
\right) = \frac{1}{\mathscr{D}} \left(
\begin{array}{c}
S_{wzz} S_{wxy} - S_{wxz} S_{wzy}\\
S_{wxx} S_{wzy} - S_{wxz} S_{wxy}\\
\end{array}
\right) \, \, \, ,
\end{equation}
where the determinant $\mathscr{D}=S_{wxx} S_{wzz} - S_{wxz}^2$. Finally, the standard errors of the parameters of the fit are obtained from the equations:
\begin{equation} \label{eq:EQA048}
(\delta \hat{p}_{1x})^2 = \sigma^2 \frac{S_{wzz}}{\mathscr{D}}
\end{equation}
and
\begin{equation} \label{eq:EQA049}
(\delta \hat{p}_{1z})^2 = \sigma^2 \frac{S_{wxx}}{\mathscr{D}} \, \, \, ,
\end{equation}
where ${\rm NDF}=N - 2$, as the linear model now contains two parameters. Regarding the expectation value of a new prediction $y_p$ at $x=x_p$ and $z=z_p$, as well as its standard error $\delta y_p$, 
Eq.~(\ref{eq:EQA043}) applies without the first term on the rhs (i.e., $\hat{p}_0$), whereas Eq.~(\ref{eq:EQA044}) now reads:
\begin{equation} \label{eq:EQA050}
(\delta y_p)^2 = \sigma^2 \left( w_p^{-1} + \frac{x_p^2 S_{wzz} + z_p^2 S_{wxx} - 2 x_p z_p S_{wxz}}{\mathscr{D}} \right) \, \, \, .
\end{equation}

\clearpage
\newpage
\section{\label{App:AppB}Numerical minimisation}

The numerical minimisation can be achieved by means of a plethora of commercial and non-commercial products. In my projects, I have always relied on the MINUIT software package \cite{James} of the CERN 
library (FORTRAN and C++ versions). In this work, the FORTRAN version was used in order that the analytical solutions, obtained after the C++ implementation of nearly all methods of Appendix \ref{App:AppA}, 
be validated. The treatment of the two methods of Appendix \ref{App:AppA4}, which feature parameter-dependent weights $w_i (p_1)$, was exclusively covered by MINUIT. (The same goes for the fit leading to 
Fig.~\ref{fig:LeavittCepheids}.) Regarding my MINUIT-based applications, each optimisation is achieved by following the sequence: SIMPLEX, MINIMIZE, MIGRAD, and MINOS. The calls to the last two methods 
involve the high-level strategy of the numerical minimisation.
\begin{itemize}
\item SIMPLEX uses the simplex method of Nelder and Mead \cite{Nelder1965}.
\item MINIMIZE calls MIGRAD, but reverts to SIMPLEX if MIGRAD fails to converge.
\item MIGRAD, undoubtedly the warhorse of the MINUIT software package, is a variable-metric method, based on the Davidon-Fletcher-Powell algorithm. The method checks for the positive-definiteness of the 
Hessian matrix.
\item MINOS carries out a detailed error analysis (separately for each parameter), taking into account all correlations among the model parameters.
\end{itemize}
All aforementioned methods admit one optional argument, fixing the maximal number of calls to each method. If this limit is reached, the corresponding method is terminated (by MINUIT, internally) regardless 
of whether or not that method had converged. To have confidence in the results of the optimisation, it is recommended to the user to always inspect the (copious) MINUIT output, so that the convergence of the 
methods and the successful termination of the application be ascertained, or any setback during the execution of the application (e.g., lack of convergence of some MINUIT methods, erroneous evaluation of the 
Hessian matrix, large estimated vertical distance to the minimum, etc.) be discovered and resolved.

Although MINUIT is robust, some attention is required in order to avoid cases where the application might get trapped in the `valley' of a local minimum. This is not unlikely to happen in non-linear problems, 
e.g., when using the two methods of Section \ref{App:AppA4} of this work, featuring parameter-dependent weights. There are several ways to safeguard against this possibility. For instance, one may repeat the 
optimisation using different initial guesses for the parameter values (thus modifying the point from where MINUIT starts the minimisation of the user-defined function), broaden the initial guesses for the 
uncertainties of the parameters of the fit, or simply restrict the parameter space by introducing lower and/or upper bounds to some or to all parameters. To cope with this issue, the MINUIT developers also 
offer the (time-consuming) method SCAN, which scans the parameter space for an optimal starting point in the minimisation problem.

Finally, it must be borne in mind that the MINOS uncertainties for the parameters of the fit do \emph{not} contain the Birge factor of Eq.~(\ref{eq:EQA026}). Consequently, it is up to the user to programme 
the user-defined function in such a way that the Birge factor be applied to the uncertainties (e.g., if ${\rm BF} > 1$), or be omitted (e.g., if ${\rm BF} < 1$ while the PDG recommendation is followed). 
Similarly, it is the user's responsibility to obtain and apply the t-multiplier to the MINOS uncertainties.

\clearpage
\newpage
\section{\label{App:AppC}On the application of the Birge factor to the uncertainties of the parameters of a fit}

When no suspicion can be cast on the modelling in a problem, and the uncertainties of the input datapoints are meaningful (see beginning of Appendix \ref{App:AppA}), the minimisation function is expected to 
follow the $\chi^2$ distribution with $N - N_p$ DoFs ($N_p$ being the number of parameters used in the fit). As a result, the expectation value of the minimisation function at the minimum should be equal to 
NDF, which implies that the expectation value of the reduced $\chi^2$, i.e., of the ratio $\sigma^2 \coloneqq \chi^2_{\rm min}/{\rm NDF}$, to be identified with the square of the Birge factor BF, is equal 
to $1$. This property of $\chi^2$-distributed quantities has led most physicists to the subliminal use of the approximate criterion $\chi^2_{\rm min}/{\rm NDF} \approx 1$ as a `rule of thumb' when assessing 
the quality of fits.

Glossing over the arbitrariness in the interpretation of an approximate equality, this practice could be meaningful in case of large samples; however, it is misleading in case of small ones. Nevertheless, 
even for large samples, the $\chi^2_{\rm min}/{\rm NDF} \approx 1$ criterion might provide a general impression about the quality of a fit, yet it represents no rigorous measure of that quality. For instance, 
does a resulting $\chi^2_{\rm min}$ value of $900$ for $800$ DoFs, yielding a $\chi^2_{\rm min}/{\rm NDF}$ value of $1.125$, indicate a satisfactory or an unsatisfactory fit at the $1$~\% significance level? 
Although one might be tempted to consider such a fit `satisfactory', the p-value, which is associated with the aforementioned numbers, is about $7.77 \cdot 10^{-3}$, i.e., below the $1$~\% significance level 
of this work, suggesting that the fit is (in the strict statistical sense) anything but satisfactory.

The correct assessment of the quality of fits rests upon the use of the p-value which is associated with the $\chi^2$ distribution for the given NDF. To this end, a number of software implementations of 
dedicated algorithms are available, e.g., see Refs.~\cite{Abramowitz1972} (Chapter on `Gamma Function and Related Functions') and \cite{Press2007}, the routine PROB of the FORTRAN implementation of the CERN 
software library (which, unlike most other CERNLIB routines, is available only in single-precision floating-point format), the functions CHIDIST or CHISQ.DIST.RT of Microsoft Excel, the function chi2cdf of 
MATLAB, etc.

A few words regarding the application of the Birge factor are due. For the sake of argument, let me assume that the $\chi^2_{\rm min}/{\rm NDF}$ result of a fit exceeds $1$ (equivalently, ${\rm BF}>1$). So 
far as physical observations, accompanied by measurement uncertainties of the input datapoints, are concerned, the quantity BF may be interpreted as the amount by which the input uncertainties need to be 
rescaled, so that the new reduced $\chi^2_{\rm min}$ becomes equal to its expectation value of $1$. If, on the contrary, the fit is already `satisfactory' (i.e., when the ratio $\chi^2_{\rm min}/{\rm NDF}$ 
does not exceed $1$), the PDG maintains that the rescale of the uncertainties of the input datapoints is not called for.

\end{document}